\documentclass[fleqn,usenatbib]{mnras}
\usepackage{newtxtext,newtxmath}
\usepackage[T1]{fontenc}

\DeclareRobustCommand{\VAN}[3]{#2}
\let\VANthebibliography\thebibliography
\def\thebibliography{\DeclareRobustCommand{\VAN}[3]{##3}\VANthebibliography}

\usepackage{graphicx}
\usepackage{dcolumn}
\usepackage{comment}
\usepackage{bm}
\usepackage{hyperref}
\usepackage{float}
\usepackage{amsmath}
\usepackage{mathtools}
\usepackage{CJK}

\newcommand{\ceffsq}{c_{\mathrm{eff}}^2}
\newcommand{\vesc}{v_{\mathrm{esc}}}

\newcommand{\texp}{t_{\mathrm{e}}}
\newcommand{\tcool}{t_{\mathrm{c}}}
\newcommand{\Msun}{M_\odot}
\newcommand{\md}{\mathrm{d}}
\newcommand{\dv}[2]{\frac{\md#1}{\md#2}}
\newcommand{\pdv}[2]{\frac{\partial#1}{\partial#2}}

\title[CR Driven Galactic Winds from the WIM]{Cosmic-Ray Driven Galactic Winds from the Warm Interstellar Medium}
\author[Modak et al.]{
Shaunak Modak,$^{1}$\thanks{E-mail: shaunakmodak@princeton.edu}
Eliot Quataert,$^{1}$
Yan-Fei Jiang (姜燕飞),$^{2}$
and Todd A. Thompson$^{3,\,4}$
\\
$^{1}$Department of Astrophysical Sciences, Princeton University, Princeton, NJ 08544, USA\\
$^{2}$Center for Computational Astrophysics, Flatiron Institute, 162 Fifth Avenue, New York, NY 10010, USA\\
$^{3}$Department of Astronomy, The Ohio State University, 140 West 18th Avenue, Columbus, OH 43210, USA\\
$^{4}$Center for Cosmology and Astro-Particle Physics (CCAPP), The Ohio State University, 191 West Woodruff Ave., Columbus, OH 43210, USA
}

\date{Accepted XXX. Received YYY; in original form ZZZ}
\pubyear{2023}

\begin{document}

\begin{CJK}{UTF8}{gbsn}

\label{firstpage}
\pagerange{\pageref{firstpage}--\pageref{lastpage}}
\maketitle

\end{CJK}

\graphicspath{{./}{figures/}}

\begin{abstract}
We study the properties of cosmic-ray (CR) driven galactic winds from the warm interstellar medium using idealized spherically symmetric time-dependent simulations. The key ingredients in the model are radiative cooling and CR-streaming-mediated heating of the gas. Cooling and CR heating balance near the base of the wind, but this equilibrium is thermally unstable, leading to a multiphase wind with large fluctuations in density and temperature. In most of our simulations, the heating eventually overwhelms cooling, leading to a rapid increase in temperature and a thermally-driven wind; the exception to this is in galaxies with the shallowest potentials, which produce nearly isothermal $T \approx 10^4\,$K winds driven by CR pressure. Many of the time-averaged wind solutions found here have a remarkable critical point structure, with two critical points. Scaled to real galaxies, we find mass outflow rates $\dot M$ somewhat larger than the observed star formation rate in low mass galaxies, and an approximately ``energy-like'' scaling $\dot M \propto v_{\rm esc}^{-2}$. The winds accelerate slowly and reach asymptotic wind speeds of only $\sim 0.4 v_{\rm esc}$. The total wind power is $\sim 1\%$ of the power from supernovae, suggesting inefficient preventive CR feedback for the physical conditions modeled here. We predict significant spatially extended emission and absorption lines from $10^4 - 10^{5.5}$\,K gas; this may correspond to extraplanar diffuse ionized gas seen in star-forming galaxies.
\end{abstract}


\section{Introduction}
Cosmic rays (CRs) are an energetically important constituent of the interstellar medium (ISM) of galaxies, and likely also the hot virialized plasma in galactic halos. CRs set the ionization state of dense gas in the ISM. They also dynamically influence the bulk of the ISM through pressure forces (mediated by the magnetic field) and potentially through heating of the thermal plasma. In the Milky Way, the pressure of CRs in the local ISM is comparable to that of the magnetic field and turbulence \citep{Boulares1990}. This has motivated a large body of work investigating whether the large CR pressure gradient in the ISM can drive a galactic wind (e.g. \citealt{Ipavich1975, Breitschwerdt1991, Everett2008, SocratesDavisRamirezRuiz}). More broadly, CR feedback is an increasingly common ingredient in models of galaxy formation (e.g. \citealt{Guo2008, Uhligh2012, Booth2013, Ruszkowski2017, Pfrommer2017, Ji2020}).

The impact of CRs on our understanding of galaxies depends in part on the poorly understood microphysics of what sets the effective mean free path of CRs in a plasma. CRs move locally at nearly the speed of light along the magnetic field, but are scattered by small-scale magnetic fluctuations. In this paper we will focus on scales larger than this scattering mean free path, in which case CR dynamics can be modeled as that of a relativistic fluid \citep{Skilling1971}. The fluctuations that scatter CRs can either be produced by an ambient turbulent cascade (``extrinsic turbulence'') or by instabilities generated by the CRs themselves (``self-confinement''). In particular, one might guess that absent any scattering by ambient turbulence, CRs would collectively stream at nearly the speed of light along the local magnetic field so as to eliminate any CR pressure gradient. However, streaming faster than the local Alfv\'en speed drives Alfv\'en waves unstable (the ``streaming instability''), which then act to scatter the CRs and limit the streaming speed to be of order the Alfv\'en speed \citep{Kulsrud1969, Bai2019}. Self-confinement theory is on somewhat firmer theoretical ground for CRs with energies $\lesssim 100$ GeV (e.g. \citealt{Blasi2012}). These CRs dominate the total CR energy density, and thus dominate the dynamical impact of CRs on the gas in galaxies. For this reason, in this work we will assume that CR transport is mediated by the streaming instability. However, we stress that neither extrinsic turbulence nor self-confinement theory fare particularly well when compared to detailed observations of CRs in the Milky Way \citep{Kempski2022, Hopkins2022}.

A key feature of self-confinement theory is that the waves generated by the CRs damp by interaction with the thermal plasma, thus transferring energy from the CRs to the thermal plasma at a (local) rate of $|{\bf v_A \cdot \nabla} p_c|$ \citep{Wentzel1971}, where $\mathbf{v_A}$ is the Alfv\'en velocity and $\nabla p_c$ is the CR pressure gradient. This heating can be energetically important in galactic winds or in lower density plasmas such as the warm ISM, hot ISM, and the intracluster medium \citep{Guo2008, Svenja2017, Kempski_thermal_instability}. \citet{Wiener_Zweibel_2013} argued that there is indirect evidence for CR heating of the warm ISM of the Milky Way in line ratios that deviate from those expected in photoionization equilibrium. Moreover, the deviations increase with increasing height above the midplane, suggesting that CR heating becomes increasingly important above the disk, where a galactic wind would originate.

Previous work on galactic winds driven by CRs has highlighted two key mechanisms by which CRs contribute to driving the winds. The first is the CR pressure gradient, and the second is CR heating of the thermal plasma, which can contribute to a thermally driven wind \citep{Ipavich1975}. The relative importance of these two CR-mediated driving mechanisms depends primarily on the rate of gas cooling. If the gas rapidly radiates away the energy supplied by the streaming CRs, then the dominant effect of the CRs is via their pressure forces. On the other hand, if cooling is inefficient, then CRs also contribute to driving the outflow by heating the thermal plasma. Despite the importance of radiative cooling in the thermodynamics of galactic winds with streaming CRs, most idealized cosmic-ray driven wind calculations either assume the gas is isothermal (the rapid cooling limit, as studied in \citealt{Mao_Ostriker} and \citealt{Quataert_streaming}) or neglect radiative cooling entirely. In the hot ISM, including CR heating but neglecting cooling can be a good approximation (e.g. \citealt{Everett2008}), but in winds driven from the warm ISM, cooling is particularly important to incorporate. Indeed, as demonstrated in \cite{HuangDavisLaunching} and \cite{HuangDavisThermal}, the inclusion of cooling can significantly alter the structure of the wind.   

There is a large literature on CR-driven winds with diffusive transport in a realistic cosmological context (e.g., \citealt{Girichidi2016,Jacob2018,Chan2019,Rathjen2021,Simpson2023}).   These calculations include many physical ingredients relevant to the formation of galactic winds, including radiative cooling, multiphase gas, and multiple stellar feedback channels.   However, diffusive CR transport does {\em not} lead to CR heating of the gas and is thus physically very different from CR transport via streaming.  Only recently have numerical methods been developed that accurately and efficiently model streaming transport in time dependent simulations \citep{Jiang_Oh_Transport,Chan2019,Thomas2021}.  As a result, the interplay of gas heating via CR streaming and cooling has not been explored in much detail.  This paper aims to bridge this gap by including streaming CRs and radiative cooling in idealized models of CR-driven galactic winds.  We hope that the insights developed here will be valuable in  interpreting CR feedback observationally and modeling it in more realistic cosmological calculations.   This work builds on recent studies carried out by a subset of the authors \citep{Quataert_streaming, Quataert_diffusion} by explicitly including radiative cooling and CR-streaming heating of the gas, rather than assuming an isothermal equation of state.  The isothermal equation of state precludes the possibility of runaway heating or cooling of the gas or thermal instability, both of which we will show are important for galactic winds driven by streaming CRs.

This paper is organized as follows. In \S\ref{sec:analytics}, we use steady-state wind theory to derive expectations for the role of cooling and CR heating in galactic winds. This includes a discussion of the very unusual critical point structure of such winds, as well as analytic approximations for the rapidly cooling, roughly isothermal base of the wind. In \S\ref{sec:numerical_sims}, we present a suite of spherically symmetric time-dependent numerical simulations of CR-driven galactic winds carried out in {\tt Athena++}, and compare their features to the steady-state predictions. In \S\ref{sec:discussion} we discuss aspects of our results including the wind mass-loss rates and terminal speeds, connections to observations, and limitations and possible generalizations of our approach. Finally, we summarize our main results in \S\ref{sec:summary}.

\section{Analytic Expectations}
\label{sec:analytics}

We begin by reviewing some of the analytic expectations for steady-state wind solutions with CR streaming and radiative cooling. In particular, we highlight the unusual critical point structure possible in such solutions, and the role of thermal instability near the base of the wind where CR heating of the gas and radiative cooling are both important.

\subsection{Steady-State Equations of Motion}

Approximating the flow as spherically symmetric and steady, conservation of mass for the wind leads to a constant mass outflow rate,
\begin{equation}
    \label{eq:Mdot}
    \dot{M} = 4\pi r^2 \rho v = \textrm{constant}
\end{equation}
which may be used to eliminate either the gas density $\rho$ or outflow speed $v$ from the equations of motion. The steady-state momentum equation for the gas is
\begin{equation}
    \label{eq:momentum}
    \rho v\dv{v}{r} = -\dv{p}{r} - \dv{p_c}{r} + \rho g
\end{equation}
where $p$ is the gas pressure, $p_c = E_c/3$ is the CR pressure, $E_c$ is the CR energy density, and $g = -\md\phi/\md r$ is the acceleration due to the galaxy's gravitational potential $\phi$.
 
Throughout this paper we assume that CR transport is regulated by the streaming instability, as is plausible for the GeV CRs that dominate the total CR energy density (e.g., \citealt{Blasi2012,Kempski2022} and references therein). In steady-state, the CR energy density $E_c$ then evolves as
\begin{equation}
    \label{eq:CR_energy}
    \nabla\cdot\textbf F_c = (\textbf v + \textbf v_s)\cdot \nabla p_c
\end{equation}
where $\textbf F_c = (4/3) E_c (\textbf v + \textbf v_s)$ is the steady-state CR energy flux, $\textbf v_s = -\textrm{sgn}(\textbf v_A \cdot \nabla p_c)\textbf v_A$ is the CR streaming velocity down the pressure gradient, and $\textbf v_A = \textbf{B}/\sqrt{4\pi\rho}$ is the Alfv\'en velocity. Note that in  equation \ref{eq:CR_energy} we have neglected both sources and sinks of CRs. In particular, we have neglected pionic losses, because they are not significant in the Milky-Way-like galaxies modeled here, though they can be important in higher-density star-forming galaxies \citep{Lacki2010}. CRs stream at exactly the Alfv\'en speed  in equation \ref{eq:CR_energy} only if the scattering rate due to waves excited by the streaming instability is very large, so that the CRs are pinned to move at exactly the speed of the waves scattering them.  In general, when the scattering rate is finite, there is a correction to pure streaming transport whose magnitude depends on the  saturation amplitude of the streaming instability (e.g., \citealt{Skilling1971,Bai2022}).  This correction is often modeled as an additional diffusion term in the CR flux $\textbf F_c$  of the form  $-\kappa \hat b (\hat b \cdot \nabla) E_c$ (where $\kappa$ is the diffusion coefficient and $\hat b$ is the direction of the local magnetic field). However, the magnitude of this correction, and indeed even whether it is actually diffusive, is not well understood and likely depends sensitively on gas temperature and density (e.g., \citealt{Wiener2018,Kempski2022}). For this reason we focus in this paper on the idealized problem of pure CR streaming with no diffusive correction.


Assuming spherical symmetry with $\md p_c/\md r < 0$,  equations \ref{eq:Mdot} and \ref{eq:CR_energy} combine to yield
\begin{equation}
    \label{eq:dpcdr}
    \dv{p_c}{r} = \frac{4}{3}\left(\frac{v + v_A/2}{v + v_A}\right)\frac{p_c}{\rho}\dv{\rho}{r} \equiv \ceffsq \dv{\rho}{r}
\end{equation}
where we follow \cite{Quataert_streaming} in defining an effective CR sound speed $\ceffsq$.

The steady-state energy equation for the gas takes the form
\begin{equation}
    \label{eq:gas_energy}
    \rho v T \dv{s}{r} = q_A - q_r
\end{equation}
Here, $s = (k/m)\log(p/\rho^\gamma)/(\gamma-1)$ is the specific entropy of the gas, $\gamma = 5/3$ is its adiabatic index, and $m$ is each gas particle's mass. The gas undergoes heating by the Alfv\'en waves generated by CR streaming at a rate
\begin{equation}
    q_A = -v_A\dv{p_c}{r}
\end{equation}
and radiates away energy at a rate
\begin{equation}
    q_r = n^2 \Lambda = \frac{\rho^2}{m^2}\Lambda
\end{equation}
for a radiative cooling function $\Lambda = \Lambda(T)$. In practice, we will use a cooling curve appropriate for collisional ionization equilibrium (CIE); see \S\ref{sec:base} and \S\ref{sec:sim_setup} for additional details. Note that we neglect photoheating of the gas. This is important near the disk of the galaxy, but CR heating scales much more weakly with density ($\propto \rho^{1/6}$ when $v \ll v_A$) than photoheating ($\propto \rho$), and so CR heating becomes increasingly dominant further from the galactic disk \citep{Wiener_Zweibel_2013}. Including photoheating is likely to only change the wind solution mildly near $T \sim 10^{4}$ K, and not at all at higher temperatures.

The resulting total energy equation for both the gas and the CRs is then
\begin{equation}
\label{eq:Edot}
    \frac{1}{r^2}\dv{}{r}\left(\dot{M}\left(\frac{1}{2}v^2 + \phi + \frac{\gamma}{\gamma-1}\frac{p}{\rho}\right) + 16\pi r^2p_c(v+v_A)\right) = -4\pi n^2 \Lambda
\end{equation}
From left to right, the terms in equation \ref{eq:Edot} include the gas kinetic energy flux $\dot{E}_k = (1/2)\dot{M}v^2$, the gravitational energy flux $\dot{E}_g = \dot{M}\phi$, the gas enthalpy flux $\dot{E}_h = \gamma\dot{M}p/((\gamma-1)\rho)$, and the (steady-state, neglecting diffusion) CR energy flux $\dot{E}_c = 4\pi r^2F_c$ including both advection of CR energy and streaming. Note that the CR heating term present in the gas energy equation \ref{eq:gas_energy} is exactly canceled by a corresponding loss term in the CR energy equation \ref{eq:CR_energy}; the total energy in the CR and gas system thus only decreases due to cooling, as is apparent on the right hand side of equation \ref{eq:Edot}.

The radial velocity, temperature, and CR pressure gradients can be found by using equations \ref{eq:dpcdr} and \ref{eq:gas_energy} to relate the CR pressure and gas pressure gradients respectively to the density gradient, and then equation \ref{eq:Mdot} to relate the density gradient to the velocity gradient. The dynamics is then specified by a coupled system of ordinary differential equations in the variables $(v, T, p_c)$. The resulting steady-state wind equation is given by
\begin{align}
    \label{eq:v_profile}
    \dv{\log v}{\log r} & = 2\frac{\big[(1 - (\gamma-1)\frac{v_A}{v})\ceffsq + \gamma a^2\big] + \frac{rg}{2} + \frac{2\pi(\gamma-1)r^3q_r}{\dot{M}}}{v^2 - \big[\ceffsq(1 - (\gamma-1)\frac{v_A}{v}) + \gamma a^2\big]} \nonumber \\
    & \equiv 2\frac{v_d^2 - v_n^2}{v^2 - v_d^2}
\end{align}
where $a^2 \equiv p/\rho = kT/m$ is the isothermal gas sound speed in the absence of CRs. 
We have defined characteristic speeds in the denominator and numerator of the wind equation \ref{eq:v_profile} as follows:
\begin{equation}
    v_d^2 \equiv \left(1 - (\gamma-1)\frac{v_A}{v}\right)\ceffsq + \gamma a^2 
    \label{eq:vd}
\end{equation}
\begin{equation}
    v_n^2 \equiv -\frac{rg}{2} - \frac{2\pi(\gamma-1)r^3q_r}{\dot{M}}
    \label{eq:vn}
\end{equation}

With these definitions, the wind will have critical points wherever $v^2 = v_d^2 = v_n^2$; we will discuss the nature of these points in greater detail in \S\ref{sec:critical}. Equation \ref{eq:v_profile} is equivalent to the wind equation in \citet{Ipavich1975} except for the addition of the cooling term in the numerator. In many problems (such as the Parker solar wind, \citealt{Parker_wind}), the speed in the denominator of the wind equation $v_d$ is the sound speed of the gas and the speed in the numerator $v_n$  is set by the sound speed and the escape speed.   This is not guaranteed, however (e.g., \citealt{Lamers1999}), and indeed is not always the case in the present problem. The total gas sound speed defined by $\mathrm{d}(p+p_c)/\mathrm{d}\rho$ is not always the same as the critical speed $v_d^2$ that appears in the wind equation due to the presence of cooling. Combining equations \ref{eq:Mdot}, \ref{eq:momentum}, and \ref{eq:v_profile} yields
\begin{equation}
    c_s^2\equiv \frac{\mathrm{d}(p+p_c)}{\mathrm{d}\rho} = v_d^2 + \frac{2\pi(\gamma-1)r^3q_r}{\dot{M}} \frac{v^2-v_d^2}{v^2-v_n^2}
    \label{eq:sound_speed}
\end{equation}

\subsection{The Role of Cooling Near the Base of the Wind}
\label{sec:cooling}

Any quasi-steady, transonic, CR-driven wind that satisfies equation \ref{eq:v_profile} is expected to initially have small velocities before accelerating outward and passing through critical point(s) at which $v^2 = v_d^2 = v_n^2$. We discuss these critical points in greater detail in \S\ref{sec:critical}. In the absence of radiative cooling, however, the critical point condition for CR driven winds with streaming and CR heating is unusual, and implies significant constraints on the properties of the flow at the base. In particular, note that $v_d^2 < 0$ if 
\begin{equation}
    \frac{v}{v_A} < \frac{(\gamma-1) \ceffsq}{\ceffsq + \gamma a^2} \sim \mathcal{O}\left(\frac{p_c}{p_c + p}\right)
    \label{eq:vdneg}
\end{equation}
where in the second expression we have used the fact that $\ceffsq$ is positive definite and of order $p_c/\rho$, as defined in equation \ref{eq:dpcdr}. \cite{Ipavich1975} noticed the fact that the critical point speed in the critical point equation could be imaginary, and attributed it to the likely existence of instabilities. Indeed, CR streaming is known to produce several distinct linear instabilities of sound waves (e.g. \citealt{Begelman1994, Quataert_streaming}).

To further explore the consequence of equation \ref{eq:vdneg}, note that if cooling is negligible, we have $v_n^2 > 0$, so equation \ref{eq:v_profile} implies $\mathrm{d}\log v/\mathrm{d}\log r < 0$ whenever $v_d^2 < 0$, i.e., the solution decelerates. Thus $v_d^2 < 0$ is incompatible with a transonic wind that accelerates outwards. Requiring $v_d^2 > 0$ at all radii implies that even at the ``base'' of the wind at small radii the outflow speed must be comparable to $v_A$ if $p_c \sim p$. This is indeed assumed to be the case in the original CR-driven wind solutions presented by \cite{Ipavich1975}. It is unclear whether such solutions could be physically extended to smaller galactic radii, where $v/v_A$ should decrease because of higher gas densities.

A simple understanding of the difficulty in realizing a highly sub-Alfv\'enic CR-driven wind absent radiative cooling can be obtained by assuming $v \ll v_A$ and assessing the consequences. In this limit, from equation \ref{eq:dpcdr}, $p_c \propto \rho^{2/3}$ and the steady-state gas energy equation reduces to $\md p/\md r = \gamma p \mathrm{d}\log \rho/\mathrm{d}r - (\gamma-1)(v_A/v) \mathrm{d}p_c/\mathrm{d}r$.  Substituting this result for the gas pressure gradient into the momentum equation and assuming hydrostatic equilbrium (consistent with the low velocities) yields
\begin{equation}
    \left[\gamma p + (\gamma-1)\left(1 - (\gamma-1)\frac{v_A}{v}\right)p_c\right] \frac{\mathrm{d} \log \rho}{\mathrm{d}r} = \rho g
    \label{eq:HElowv}
\end{equation}
where we have eliminated $\md p_c/\md r$ in favor of $\md\rho/\md r$ using $p_c\propto\rho^{2/3}$. If $v \ll v_A$, however, equation \ref{eq:HElowv} implies $\mathrm{d}\rho/\mathrm{d}r > 0$ and thus $\mathrm{d}p_c/\mathrm{d}r > 0$, which is inconsistent with the assumption that the CRs stream outwards. Physically, the issue is that without radiative cooling, the CR heating of the gas ($-v_A \mathrm{d}p_c/\mathrm{d}r$) is too large to realize a quasi-hydrostatic solution if $v_A \gg v$.

As we will show in this work, radiative cooling can remove the difficulties we have just highlighted in obtaining sub-Alfv\'enic wind solutions. In particular, its presence allows for the possibility that $v_n^2 < 0$, so accelerating winds are once again attainable. A strong indication of this lies in the existence of isothermal CR-driven winds with $v \ll v_A$ (e.g. \citealt{Mao_Ostriker, Quataert_streaming}); these effectively represent the limit of very strong cooling regulating the gas temperature. We discuss the formal isothermal limit of the steady-state equations considered here in more detail in \S\ref{sec:iso}.

\subsection{Critical Points}
\label{sec:critical}

At a critical point of the wind, both $v^2 = v_n^2$ and $v^2 = v_d^2$. From the ``numerator'' equation, we arrive at a quartic equation for the velocity at a critical point $v_c$ in terms of the position $r_c$ and temperature $T_c$ at the critical point:
\begin{equation}
    \label{eq:wind_num_crit}
    v_c^4 + \frac{r_cg(r_c)}{2}v_c^2 + \frac{\dot{M}}{8\pi m^2 r_c}(\gamma-1)\Lambda(T_c) = 0
\end{equation}
The wind thus has critical points when its velocity is
\begin{equation}
    \label{eq:crit_vel}
    v_{c,\,\pm} = \frac{\sqrt{-r_c g(r_c)}}{2}\left(1\pm \left(1-\frac{8(\gamma-1)\texp{}}{\tcool{}}\right)^{1/2}\right)^{1/2}
\end{equation}
where we have identified the cooling time $\tcool = p/q_r$ and some effective expansion time $\texp = va^2/(rg^2)$, which should be evaluated at the critical point $r_c$. Unlike the critical points of the isothermal problem studied in \cite{Mao_Ostriker,Quataert_streaming}, at each radius, two critical speeds are possible in general, as long as $8(\gamma-1)\texp < \tcool$, i.e. when cooling is not so rapid that the gas remains isothermal but is significant enough that we cannot neglect the final term of equation \ref{eq:wind_num_crit}. Note that taking appropriate limits, $v_{c,\,+}$ is the unique critical speed in both the isothermal case (when $(\gamma-1)\Lambda = 0$, see \S\ref{sec:iso}) and when cooling is negligible (when $\Lambda \equiv 0$).

In a typical transonic wind, there can only be an odd number of critical points, because the wind speed is initially below the local sound speed, but must exceed the local sound speed as $r\to\infty$: any intermediate regions in which $c^2_s(r) > v(r)^2$ must be followed by an additional crossing at which $v(r)^2 > c^2_s(r)$ (e.g. \citealt{Lamers1999}). Indeed, previous hydrodynamic CR-driven winds that we are aware of all pass through only one critical point (e.g. \citealt{Ipavich1975, Breitschwerdt1991, Everett2008, Mao_Ostriker}).

Remarkably, we will see that the time average of many of the time-dependent simulations presented in this work pass through two critical points. This is possible because $v_d^2 < 0$ and $v_n^2 < 0$ near the base of the wind where $v \ll v_A$ (see equations \ref{eq:v_profile} and \ref{eq:vd}). Thus the solution starts ``supersonic'' in the sense that $v^2 > 0 > v_d^2$, transitions to ``subsonic'' at a first critical point, and then at a larger radius undergoes a more conventional subsonic to supersonic transition at a second critical point. At the first critical point, radiative cooling is energetically important and $v_n^2 < 0$, while the second critical point is essentially the classic Parker critical point \citep{Parker_wind}.

It would be very reasonable to doubt that steady-state solutions with the critical point structure suggested here could be realized, given the likely instabilities implied by $v_d^2 < 0$. However, we note that the isothermal calculations presented in \cite{Quataert_streaming} are unstable, and yet present the expected isothermal critical point structure. Indeed, the time-dependent solutions we present in \S\ref{sec:numerical_sims} are unstable. As we shall see in \S\ref{sec:critical_sim}, though, the time-averaged solutions nonetheless have the unusual critical point structure suggested by the steady-state equations.

\subsection{The Isothermal Limit}
\label{sec:iso}
It is instructive to consider how the steady-state equations derived here reduce to the corresponding isothermal equations used in previous work (e.g. \citealt{Mao_Ostriker, Quataert_streaming}). There are two ways to take the isothermal limit of our equations. One is to take $\Lambda\to\infty$, i.e. the gas rapidly radiates away all added heat to maintain a fixed temperature, and also set $\gamma\to 1$, since the equation of state becomes $p = \rho a^2$ for fixed sound speed $a$. Therefore, we must carefully consider the behavior of the combination $(\gamma-1)\Lambda$: rearranging the expression for the gas energy in equation \ref{eq:gas_energy}, we find
\begin{equation}
    \label{eq:isothermal_limit}
    (\gamma - 1)\Lambda = \frac{pv}{n^2}\left(-\dv{\log T}{r} + (\gamma-1)\left(1 - \frac{v_A}{v}\frac{\ceffsq}{a^2}\right)\dv{\log\rho}{r}\right)
\end{equation}
As we take the isothermal limit, the first term on the right tends to zero, since $T$ is constant, and since $\md\log\rho/\md r$ must remain finite, the second term is also zero as $\gamma\to 1$. So, comparing any steady-state expression here to the analogous result in the isothermal case can be done by setting $\gamma=1$ and the combination $(\gamma-1)\Lambda = 0$. For example, doing so for equation \ref{eq:v_profile} reproduces the isothermal wind equation studied in \cite{Mao_Ostriker} and \cite{Quataert_streaming}. In particular, the isothermal limit of equation \ref{eq:v_profile} corresponds to $v_n^2 = -rg/2$ and $v_d^2 = \ceffsq + a^2$. Therefore, $v_d^2 > 0$ and $v_n^2 > 0$ as well, so none of the difficulties with $v \ll v_A$ highlighted in \S\ref{sec:cooling} are present in the isothermal limit when cooling is rapid.

A second way to consider the isothermal limit is to calculate the cooling needed to maintain an exactly constant temperature for $\gamma \neq 1$. In that case, from equation \ref{eq:isothermal_limit}, we see that an isothermal profile is possible only if
\begin{equation}
    \label{eq:isothermal_balance}
    n^2 \Lambda = (va^2 - v_A\ceffsq)\dv{\rho}{r}
\end{equation}
Substituting this result into the gas energy equation \ref{eq:gas_energy} to eliminate $\Lambda$ again reproduces the isothermal wind equation.

\subsection{The Hydrostatic, Isothermal Base of the Wind}
\label{sec:base}

Near the base of the wind, where the densities are highest, radiative cooling is energetically important for the winds considered in this work. For any sub-Alfv\'enic and sub-sonic wind with $v \ll v_A$ and $v \ll a$ near the base, from the gas energy equation \ref{eq:gas_energy}, the heating and cooling rates must balance:
\begin{equation}
    \label{eq:balanced_heatcool}
    v_A\dv{p_c}{r} = -n^2\Lambda
\end{equation}
Because this condition matches the criterion of equation \ref{eq:isothermal_balance} in the limit $v \ll v_A$, we expect that any transonic wind with initially small velocities will include an approximately isothermal region near the base. For equation \ref{eq:balanced_heatcool} to be realizable, however, the cooling rate $\Lambda(T)$ must be large enough. A rough estimate of the minimum required cooling rate, $\Lambda_{\rm min}$, can be found by taking $\md p_c/\md r \simeq (2/3)(p_c/\rho)\md\rho/ \md r$ in equation \ref{eq:balanced_heatcool} (because $v \ll v_A$) and using the hydrostatic, isothermal approximation for the density gradient derived in equation \ref{eq:hydrostatic_eq} below:
\begin{align}
    \label{eq:lambda_min}
    \Lambda > \Lambda_{\mathrm{min}} & \equiv \frac{-2 p_c g v_A}{3 n^2(a_0^2+c_{\rm eff}^2)} \nonumber \\
    & \approx 1.5 \times10^{-26} \, \frac{\mathrm{erg}\,\mathrm{cm}^3}{\mathrm{s}}  \left(\frac{p_c}{{\mathrm{eV}/\mathrm{cm}^{3}}}\right) \left(\frac{v_A}{10\,\mathrm{km}/\mathrm{s}}\right) \nonumber \\    
    & \indent\indent\indent \times \left(\frac{r}{\rm kpc} \right)^{-1}\left(\frac{n}{\mathrm{cm}^{-3}}\right)^{-2} \left(\frac{\vesc}{\rm 20 \sqrt{a_0^2 + \ceffsq}} \right)^{2}
    \end{align}
where we have approximated $rg \simeq (1/9)v_\mathrm{esc}^2$ as is true at the base of the Hernquist models we use in the simulations presented in \S\ref{sec:numerical_sims}; see \S\ref{sec:sim_setup} for details.

In CIE, a typical value for the radiative cooling function from $T\sim 10^4 - 10^8$\,K is $\Lambda\sim 10^{-23} - 10^{-22}$\,erg\,cm$^{3}$\,s$^{-1}$ (e.g. \citealt{Draine}).

\begin{figure}
    \centering
    \includegraphics[width=0.48\textwidth]{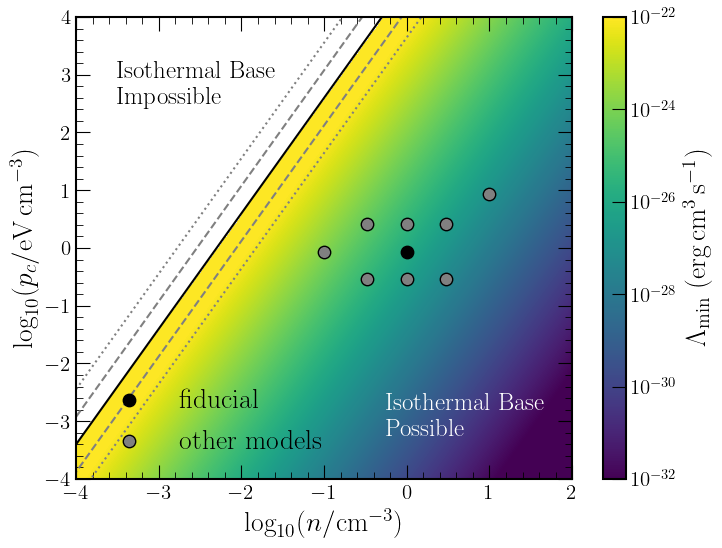}
    \caption{The minimum value of the cooling function for which cooling can balance CR heating, $\Lambda_\mathrm{min}$, as a function of the CR pressure $p_c$ and number density $n$ near the base radius for $r = 1$\,kpc, $v_A =  10$\,km/s, and $\vesc = 20\sqrt{a_0^2 + \ceffsq}$. The black solid line indicates the border between the parameter space in which we expect an initially isothermal region (colored by $\Lambda_\mathrm{min}$ value) and the region in which cooling is not strong enough to offset CR heating (in white) for the parameters listed above. The upper and lower grey dashed lines indicate how the boundary shifts if instead $v_A = 10/3$\,km/s or $30$\,km/s respectively, and the upper and lower grey dotted lines indicate how it shifts if instead $\vesc = 60\sqrt{a_0^2 + \ceffsq}$ or $(20/3)\sqrt{a_0^2 + \ceffsq}$ respectively. The black and grey points indicate values of the base density and CR pressure used in our numerical simulations presented in section \ref{sec:numerical_sims}, with the black point indicating our fiducial choice.}
    \label{fig:LambdaMin}
\end{figure}

Figure \ref{fig:LambdaMin} shows values of $\Lambda_\mathrm{min}$ as a function of $p_c$ and $n$ near the base radius, for $r = 1$\,kpc, $v_A = 10$\,km/s, and $\vesc = 20\sqrt{a_0^2 + \ceffsq}$, and indicates the region in which $\Lambda > \Lambda_\mathrm{min}$ is not achievable.

In the hot ISM, because the number density may be as low as $\sim 10^{-3}\,$cm$^{-3}$, equation \ref{eq:lambda_min} may not be satisfied, though at such high temperatures, thermal driving of the wind by the gas is likely to be comparable in importance to CRs anyway. Other than this low density regime, however, the condition in equation \ref{eq:lambda_min} is easily met across a wide range of densities, CR pressures, and magnetic field strengths appropriate for the warm ISM ($T\sim 10^4$\,K). This includes both Milky-Way-like physical conditions and those in starburst galaxies with higher gas densities and CR pressures.

When $\Lambda_\mathrm{min} \ll 10^{-23} - 10^{-22}$\,erg\,cm$^{3}\,$s$^{-1}$, gas near the base of the wind cools rapidly and remains at a roughly constant temperature of $\sim 10^4$ K. Note that this is also true if the gas is roughly in photoionization equilibrium. We can approximate the gas in this region as hydrostatic and isothermal, with constant gas sound speed $a_0$. In this approximation, the momentum equation simply yields
\begin{equation}
    \label{eq:hydrostatic_eq}
    \dv{p}{r} + \dv{p_c}{r} = (a_0^2+\ceffsq)\dv{\rho}{r} = \rho g
\end{equation}
Approximating $p_c\propto\rho^{2/3}$ because $v \ll v_A$ and integrating, we arrive at an implicit equation specifying the density profile $\rho(r)$,
\begin{equation}
    \label{eq:implicit_rho}
    a_0^2\log\left(\frac{\rho}{\rho_0}\right) + 2\frac{p_{c0}}{\rho_0}\left(1 - \left(\frac{\rho}{\rho_0}\right)^{-1/3}\right) = \phi(r_0) - \phi(r)
\end{equation}
where $\phi$ is the gravitational potential and $r_0$, $\rho_0$, and $p_{c0}$ are the base radius, density, and CR pressure respectively. This generalizes the results of \cite{Quataert_streaming} to an arbitrary gravitational potential $\phi(r)$.

Given the density profile in equation \ref{eq:implicit_rho}, we can roughly estimate the resulting temperature profile required for cooling to  balance the CR heating according to equation \ref{eq:balanced_heatcool}, namely
\begin{equation}
    \label{eq:T_approx}
    T(r) = \Lambda^{-1}\left(\frac{v_A(r)\ceffsq(r)}{n(r)^2}\left|\dv{\rho}{r}(r)\right|\right)
\end{equation}
Here, the inverse is well-defined, because the cooling curve near a typical base temperature of $T_0 = 10^4$\,K is monotonically increasing \citep{Draine}. In particular, in CIE, the cooling curve is given roughly by $\Lambda(T) = A(T - T_0)$ for a normalization constant $A \approx 1.3\times 10^{-26} \, \mathrm{erg\, cm}^3\,\mathrm{s}^{-1}\,\mathrm{K}^{-1}$. Substituting the result of equation \ref{eq:hydrostatic_eq}, we find
\begin{equation}
    \label{eq:T_approx_specifics}
    T(r) = T_0 - \frac{m v_A(r) \, g(r)}{An(r)\big[1 + a_0^2/\ceffsq(r)\big]}
\end{equation}
To be clear, the approximation leading to equation \ref{eq:T_approx_specifics} is that we first estimate the density profile assuming the gas is isothermal, and then derive an updated temperature profile for the gas using that isothermal density profile.

This also provides an estimate of the temperature profile near the base of the wind that determines roughly the extent of the isothermal region, beyond which the approximations used here may not be reliable. We characterize the extent of the isothermal region by the difference between the radius at which the temperature first exceeds $T = 1.5 \times 10^4$\,K and the base radius. Throughout the allowed parameter space where $\Lambda > \Lambda_{min}$ depicted in Figure \ref{fig:LambdaMin}, the thickness of this isothermal base varies from $\sim 10^{-2}$\,kpc at low $n$ and $p_c$ to $\sim 1$\,kpc at high $n$ and $p_c$. For most Milky-Way-like base parameters (such as those used in the simulations presented in \S\ref{sec:numerical_sims}), the typical thickness of the isothermal base ranges from $\sim 0.03 - 0.3$\,kpc. It is very likely that in more realistic models the extent of the isothermal region will be larger than in our idealized spherical calculations. In particular, in winds from a galactic disk, the isothermal region will likely be larger because the gas density decreases more slowly as gas moves away from the galaxy midplane, enhancing the importance of cooling relative to the spherical models considered here.

\begin{figure*}
    \centering
    \includegraphics[width=0.95\textwidth]{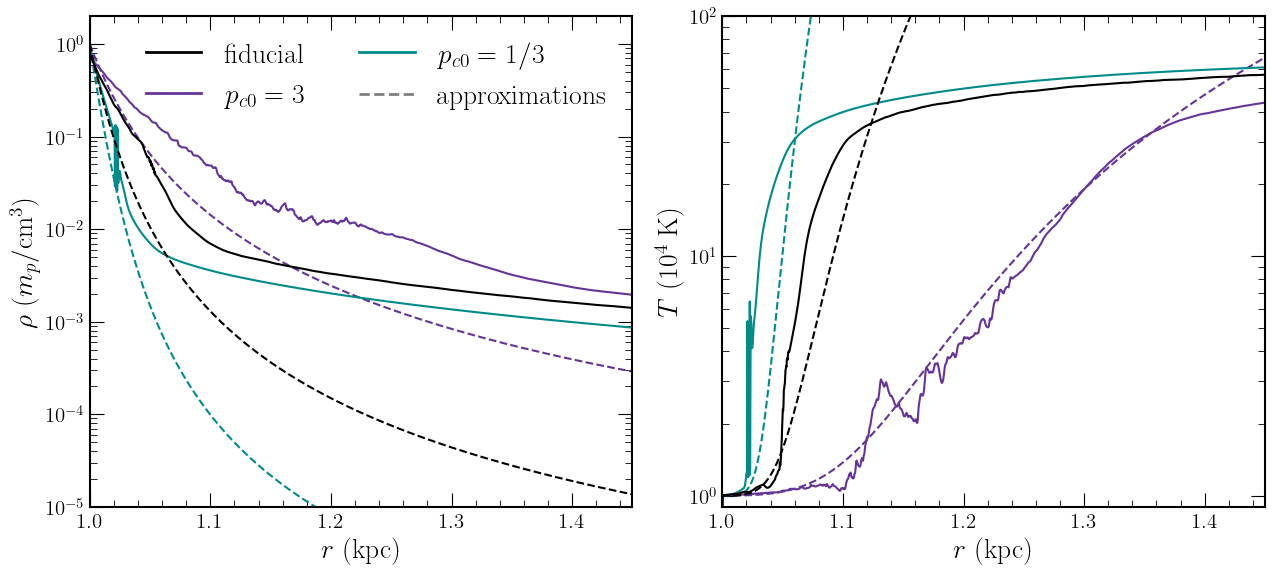}
    \caption{A comparison between the approximate density and temperature profiles for the base of the wind (equations \ref{eq:implicit_rho} and \ref{eq:T_approx}, dashed curves) and the time-averaged profiles (solid curves) from three numerical simulations, which we describe in \S\ref{sec:numerical_sims}. Specifically, the black, purple, and teal curves are simulations 1, 7, and 8 of Table \ref{tab:runs} respectively. These parameter values were chosen to demonstrate that the analytic approximations hold reasonably well regardless of whether CR pressure (purple) or gas pressure (teal) dominates near the base.}
    \label{fig:T_spike_approx}
\end{figure*}

Figure \ref{fig:T_spike_approx} compares the density and temperature profiles resulting from these approximations to two of the numerical results presented in \S\ref{sec:numerical_sims}.
To calculate the profiles plotted in the Figure, we use the same Hernquist gravitational potential and split-monopole Alfv\'en speed profiles as in the simulations; see \S\ref{sec:sim_setup} for more details. Because these approximate profiles assume hydrostatic equilibrium at a fixed temperature and rely on a linear approximation to the cooling curve valid only for temperatures below $\sim 1.2\times 10^4\,$K, we expect them to diverge from the simulations at fairly low radii. However, overall, we find reasonable agreement across a wide range of base parameters for the structure of the solution near the base of the wind: the extent of the approximately isothermal region predicted by equation \ref{eq:T_approx} is within $\sim 5$\% of that found in the simulations.

One key prediction of Figure \ref{fig:T_spike_approx} is that as the density drops with increasing distance, the temperature required to maintain a balance between heating and cooling increases, because the cooling function has to increase to compensate for the lower density. As we now discuss, this increase in temperature inevitably leads to the onset of thermal instability.

\subsection{The Onset of Thermal Instability}
\label{sec:instability_onset}

Although the wind is approximately isothermal near its base, the balance between heating and cooling that allows it to remain so can be linearly unstable. As detailed in \cite{Kempski_thermal_instability}, for a given cooling curve $\Lambda(T)$, thermal instability will set in approximately once 
\begin{equation}
    \label{eq:instability_criterion}
    \Lambda_{T} \equiv \pdv{\log\Lambda}{\log T} < \Lambda_{T,\,C}
\end{equation}
where the critical value for the logarithmic slope of the cooling curve $\Lambda_{T,\,C}$ is given by
\begin{equation}
    \label{eq:lambdaTC}
    \Lambda_{T,\,C} \simeq \frac{11}{6}\left(1 + \frac{p_c}{1.19p}\right)^{-1.13}
\end{equation}
The critical value $\Lambda_{T,\,C}$ ranges from $\Lambda_{T,\,C}=0$ for CR-dominated plasmas with $p_c \gg p$ to $\Lambda_{T,\,C}=11/6$ for gas-pressure-dominated plasmas. Because $\Lambda_{T,\,C} > 0$, in practice, thermal instability sets in when the first local maximum of $\Lambda(T)$ is reached, which occurs at approximately $T\sim 1.75\times 10^4\,\mathrm{K}$ for our cooling curve in CIE. Figure \ref{fig:T_spike_approx} shows that this temperature is reached not far from the wind base in many cases. The simulations presented in \S \ref{sec:numerical_sims} will quantify the nonlinear saturation of thermal instability for the CR-driven wind problem (in spherical symmetry; see \citealt{HuangDavisThermal} for multi-dimensional simulations).

\section{Numerical Simulations}
\label{sec:numerical_sims}

In what follows, we present time-dependent numerical solutions for CR-streaming-driven galactic winds by solving the spherically symmetric CR hydrodynamic equations. We elected to carry out time-dependent simulations rather than attempt to find steady-state solutions for several reasons. First, the known instabilities present in the case of CR streaming can significantly modify the dynamics of the wind (e.g. \citealt{Quataert_streaming, Tsung_Oh_Jiang_Staircase, HuangDavisLaunching, HuangDavisThermal}), so that capturing the time-dependent instabilities is important. Additionally, the difficulty realizing steady-state solutions with initial $v \ll v_A$ as discussed in \S \ref{sec:critical} motivates time-dependent simulations to see if the same difficulties are in fact present in the general problem. Finally, the steady-state equations with radiative cooling are extremely stiff near the base where cooling is important, so they cannot be easily directly integrated.

\subsection{Simulation Setup and Parameters}
\label{sec:sim_setup}

In our time-dependent simulations, we solve the CR hydrodynamic equations in spherical symmetry using the numerical scheme described in \cite{Jiang_Oh_Transport}, implemented in the {\tt Athena++} code \citep{Stone_athena}:
\begin{align}
    \label{eq:hydro_eqs}
    \pdv{\rho}{t} + \frac{1}{r^2}\pdv{}{r}(r^2\rho v) & = 0 \\
    \pdv{}{t}(\rho v) + \frac{1}{r^2}\pdv{}{r}(r^2\rho v^2) & = \rho g - \dv{p}{r} + \sigma_c(F_c - (E_c + p_c)v) \\
    \pdv{E}{t} + \frac{1}{r^2}\frac{\partial}{\partial r}(r^2(E + p)v) & = (v + v_s)\sigma_c(F_c - (E_c+p_c)v) - \frac{\rho^2}{m^2}\Lambda \\
    \pdv{E_c}{t} + \frac{1}{r^2}\pdv{}{r}(r^2F_c) & = -(v+v_s)\sigma_c(F_c - (E_c+p_c)v) \\
    \label{eq:CRflux}
    \frac{1}{v_m^2}\pdv{F_c}{t} + \pdv{p_c}{r} & = -\sigma_c(F_c - (E_c+p_c)v)
\end{align}
Here, $v_s = -\textrm{sgn}\left(v_A \md p_c/\md r\right)v_A$ is the CR streaming speed, and $\sigma_c^{-1} = 3\kappa + (E_c+p_c)v_A\left|\md p_c/\md r\right|^{-1}$ is the ratio between the total CR flux and the CR pressure gradient. Note that $\sigma_c$ captures the effects of both diffusion and streaming. Because we focus here on winds driven by CR streaming instead of diffusion, we take $\kappa = 10^{-3}$\,kpc\,km/s in all of our simulations; this small value is numerically useful but contributes little physically to the CR transport. The speed $v_m$ in the CR flux equation \ref{eq:CRflux} is the reduced speed of light; we take this to be $v_m = 3\times 10^4$\,km/s, chosen to be much larger than $v$ and $v_A$ throughout the simulation domain. To check that reasonable variations in $v_m$ do not alter our results, we repeat the $\vesc = 250$\,km/s simulation (row 12 of Table \ref{tab:runs}) with $v_m = 10^4$\,km/s and $v_m = 10^5$\,km/s, and find no statistical difference in the final profiles.

For the magnetic field, we consider a steady split-monopole configuration, $B(r) = B_0(r/r_0)^{-2}$ where $r_0$ is taken here to be the inner radius in the simulation (the ``base'' of the wind). Due to the spherical symmetry, the field is not separately evolved; it is only used in defining the Alfv\'en speed, which is then
\begin{equation}
    \label{eq:vA}
    v_A(r) = v_{A0}\left(\frac{r}{r_0}\right)^{-2}\left(\frac{\rho(r)}{\rho_0}\right)^{-1/2}
\end{equation}
where $v_{A0} = B_0/\sqrt{4\pi\rho_0}$ and $\rho_0$ is the base density. For the galaxy's potential, we use the Hernquist model \citep{Hernquist_potential}
\begin{equation}
    \label{eq:hernquist}
    \phi(r) = -\frac{1}{2}\frac{\vesc^2}{1 + r/b}
\end{equation}
where $b$ is a chosen scale length that we set to $b = 2r_0$, and $\vesc$ is the escape speed from $r = 0$ (note that the escape speed from the base of the wind is a factor of $\sqrt{1+r_0/b} \approx 1.22$ smaller).
For reference, the mass enclosed within a radius $r$ of the galaxy is then given by 
\begin{align}
    \label{eq:enclosed_mass}
    M(r) & = \frac{b\vesc^2}{2G}\left(1 + \frac{b}{r}\right)^{-2} \nonumber \\
    & \approx 1.45\times 10^{10}\Msun \left(\frac{b}{2\,\mathrm{kpc}}\right)\left(\frac{\vesc}{250\,\mathrm{km/s}}\right)^2 \left(1 + \frac{b}{r}\right)^{-2}
\end{align}
Simulations with the isothermal potential used in \citet{Quataert_streaming, Quataert_diffusion} gave similar results to those presented here. Because of the complexity of the critical point structure highlighted in \S\ref{sec:critical}, we chose the Hernquist potential over the isothermal potential, since the former has a well-defined escape speed.

We use a radiative cooling curve appropriate for solar abundance gas in CIE\footnote{Our results should not depend strongly on the gas metallicity, since the occurrence of thermal instability only relies on the presence of a local minimum in the cooling curve between $10^4 - 10^5$\,K, which is present from $\sim 0.01 - 2 Z_\odot$ (see e.g. Figure 34.2 of \citealt{Draine}). Metallicity gradients in the wind would therefore not strongly affect its structure, though increased metallicity at the base would lead to a larger region near the base (see \S\ref{sec:base}) due to the increased cooling rate.}, and approximate its form by fitting a piecewise power law to Figure 1 of \cite{Ji_cooling}. Below $\sim11500$\,K, it is linearly interpolated so that $\Lambda(10^4\,\textrm{K}) = 0$, and $\Lambda(T) = 0$ for all $T < 10^4$\,K as well.

Our simulation domain extends from an inner radius of 1\,kpc to an outer radius of $10$\,kpc in most runs, although we additionally simulate a larger box with an outer radius of $100$\,kpc for the fiducial set of parameters to check convergence and in any set of parameters for which the second critical point occurs outside of $10$\,kpc. The simulations are run on a radial grid of 8704 logarithmically spaced points, and we have checked that using a grid with twice the resolution does not significantly alter the resulting steady-state profiles in the fiducial and $\vesc = 250$\,km/s models. At the inner boundary, we fix $\rho_0$ and the base CR pressure $p_{c0}$, and enforce hydrostatic equilibrium, choosing $\md p_c/\md r = \md p/\md r$. In the ghost zones, the velocity is set so that $\dot{M}$ is constant between the last active zone and the ghost zones. At the outer boundary, we match the gradients of $\rho$, $p_c$, and the CR flux $F_c$ across the boundary, and again set the velocity in the ghost zones by requiring $\dot{M}$ to be constant. We initialize the gas as isothermal with $T_i(r) = 10^4$\,K, set $v_i(r) \equiv 0$, and equate the gas and CR pressures, $p_{ci} = p_i$. The initial density is specified as $\rho_i(r) = \rho_0(r/r_0)^{r_0g(r_0)/2a_i^2}$, where $a_i^2 = kT_i/m_p$ is the initial sound speed squared. We do not impose any temperature boundary conditions, only an initial condition; we conducted tests with varied $T_i$ and found that the resulting statistical steady-state did not change. The reason is that the simulations all lie in the parameter space where the cooling time at the base of the wind is sufficiently short to bring the gas to $T \sim 10^4$\,K (as expected from Figure \ref{fig:LambdaMin} and associated discussion in \S\ref{sec:base}).

\begin{table*}
    \centering
    \large
    \caption{A summary of the simulation suite, with columns including the base gas density, base CR pressure, base Alfv\'en speed, Hernquist escape velocity parameter, base radius of the simulation box, outer radius of the simulation box, mass-loss rate, wind speed at 80\% of the outer radius, theoretical maximum wind speed (according to equation \ref{eq:v_infty}) at 80\% of the outer radius, and energy flux at 80\% of the outer radius normalized by the base CR energy flux. The last three columns' quantities are evaluated at 80\% of the outer radius to avoid the possibility of spurious boundary effects. Simulation 1 is the fiducial run, and runs are referenced in the text by the parameter(s) that differs from the fiducial value (e.g. $\vesc = 250$\,km/s for simulation 12, which matches fiducial base values except for the strength of the potential). The parameters that differ are from fiducial are bolded in the Table for convenience.}
    \def\arraystretch{1.1}
    \begin{tabular}{cccccccccccc}
         & $\rho_0$ & $p_{c0}$ & $v_{A0}$ & $\vesc$ & $r_0$ & $r_\textrm{out}$ & $\dot{M}$ & $v(0.8r_\textrm{out})$ & $v_\infty(0.8r_\textrm{out})$ & $\dot{E}(0.8r_\textrm{out})$ \\
        
         ID & ($m_p/\textrm{cm}^{3}$) & ($0.86$\,eV/cm$^3$) & (km/s) & (km/s) & (kpc) & (kpc) & ($\Msun$/yr) & (km/s) & (km/s) & $\left(\dot{E}_{c0}\right)$ \\ \hline
        
        1 & 1 & 1 & 10 & 420 & 1 & 10 & 0.014 & 101 & 155 & 0.11 \\
        2 & 1 & 1 & 10 & 420 & 1 & \textbf{100} & 0.014 & 151 & 156 & 0.11 \\
        3 & \textbf{3} & \textbf{3} & 10 & 420 & 1 & 10 & 0.034 & 129 & 179 & 0.12 \\
        4 & \textbf{1/3} & \textbf{1/3} & 10 & 420 & 1 & 10 & 0.0057 & 72 & 136 & 0.09 \\ 
        5 & \textbf{1/3} & \textbf{1/3} & 10 & 420 & 1 & \textbf{100} & 0.0057 & 124 & 130 & 0.09 \\
        6 & \textbf{10} & \textbf{10} & 10 & 420 & 1 & 10 & 0.092 & 161 & 207 & 0.14 \\
        7 & 1 & \textbf{3} & 10 & 420 & 1 & 10 & 0.033 & 117 & 160 & 0.10 \\
        8 & 1 & \textbf{1/3} & 10 & 420 & 1 & 10 & 0.0044 & 77 & 149 & 0.09 \\
        9 & 1 & 1 & \textbf{30} & 420 & 1 & 10 & 0.036 & 137 & 194 & 0.18 \\
        10 & 1 & 1 & \textbf{10/3} & 420 & 1 & 10 & 0.0048 & 63 & 123 & 0.05 \\
        11 & 1 & 1 & 10 & \textbf{330} & 1 & 10 & 0.015 & 116 & 152 & 0.14 \\
        12 & 1 & 1 & 10 & \textbf{250} & 1 & 10 & 0.020 & 100 & 125 & 0.16 \\
        13 & 1 & 1 & 10 & \textbf{180} & 1 & 10 & 0.038 & 43 & 56 & 0.13 \\ 
        14 & 1 & 1 & 10 & \textbf{130} & 1 & 10 & 0.082 & 35 & 41 & 0.14 \\
        15 & \textbf{0.1} & 1 & 10 & 420 & 1 & 10 & 0.016 & 64 & 107 & 0.07 \\ 
        16 & 1 & 1 & 10 & 420 & \textbf{5} & \textbf{50} & 0.19 & 133 & 185 & 0.11 \\ \hline
    \label{tab:runs}       
    \end{tabular}
\end{table*}

Table \ref{tab:runs} gives a summary of the simulation parameter choices in physical units as well as measures of the resulting mass-loss rates, energy fluxes, and outflow speeds. Because we use a realistic atomic cooling function, our simulations necessarily use real units. Our fiducial simulation (the first row in the table) takes base conditions typical of the warm ISM in the Milky Way: $\rho_0 = 1\,m_p/\mathrm{cm}^{3}$, $p_{c0} = 0.86$\,eV/cm$^{3}$, and $v_{A0} = 10$\,km/s at $r_0$ = 1\,kpc. We take $\vesc = 420$\,km/s similar to that of a Milky-Way-like galaxy. In addition to the fiducial parameter choices described above, we consider variation in
\begin{itemize}
    \item the base gas density at fixed base CR sound speed (modifications to both $\rho_0$ and $p_{c0}$)
    \item the base ratio of CR pressure to gas pressure (modifications to $p_{c0}$ at fixed $\rho_0$)
    \item the magnetic field strength (modifications to $v_{A0}$ at fixed $\rho_0$)
    \item the galaxy escape speed from $130-420$\,km/s, to represent galaxies of different masses
    \item the launching radius of the wind, $r_0$ (at fixed $r_0/b = 2$)
\end{itemize}

We note that our $\vesc = 420$, $250$, and $130$\,km/s simulations roughly map onto the $V_g = 10$, $6$, and $3$ simulations in \cite{Quataert_streaming} respectively (taking the assumed constant sound speed = $10$\,km/s in the latter simulations). Assuming an NFW halo with a concentration parameter $c = 7$, these escape speed parameter choices correspond to virial mass values ranging from $M_{200} \approx 2 \times 10^{10}\Msun$ for the $\vesc = 130$\,km/s model to $M_{200} \approx 6\times 10^{11}\Msun$ for the $\vesc = 420$\,km/s model.

\subsection{Overview of the Simulation Results}
\label{sec:sims_overview}

\begin{figure*}
    \centering
    \includegraphics[width=0.95\textwidth]{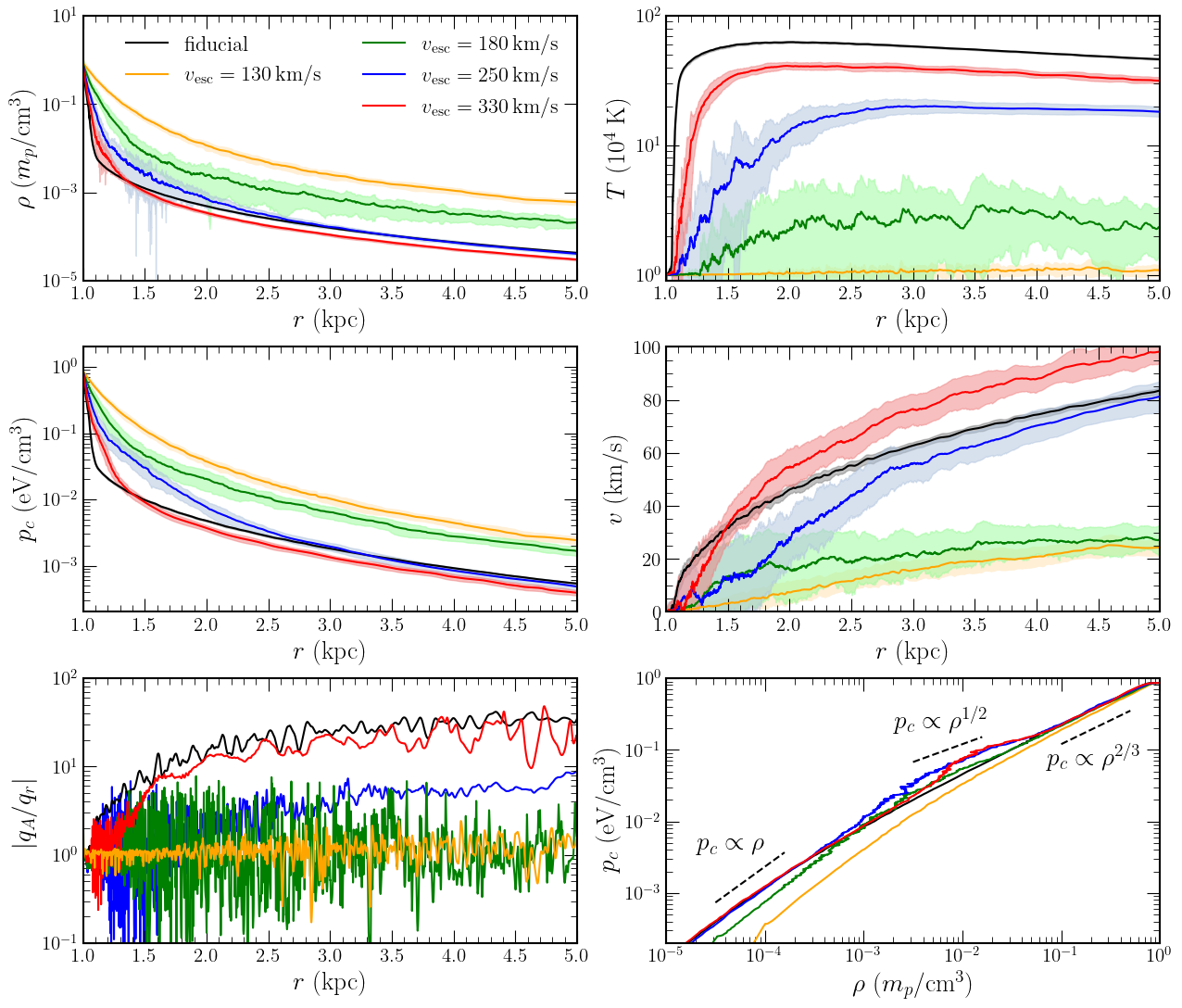}
    \caption{The steady-state radial density (upper left), temperature (upper right), CR pressure (middle left), wind speed (middle right), heating-to-cooling ratio (bottom left), and effective CR equation of state $p_c(\rho)$ (bottom right) profiles for the varied $\vesc$ simulations (simulations 1 and 11-14 of Table \ref{tab:runs}).  In the upper two rows, the shaded regions indicate $\pm 1\sigma$ (temporal) variations at each radius. In the bottom right panel, the dashed lines on the left, middle, and right of the plot indicate power law slopes of $p_c\propto\rho$, $p_c\propto\rho^{1/2}$, and $p_c\propto\rho^{2/3}$ respectively. As $\vesc$ is increased from 130\,km/s (orange) to 180 (green), 250 (blue), 330 (red), and finally the fiducial 420\,km/s (black), the extent of the strong cooling region near the base of the wind decreases. The decrease in cooling at higher $\vesc$ leads to progressively sharper spikes in temperature close to the base of the wind due to CR heating overwhelming cooling. Intermediate $\vesc$ solutions are the most time variable due to thermal instability, apparent here as larger $\pm 1\sigma$ variations and radial fluctuations even in the time-averaged profiles.}
    \label{fig:summary_profiles}
\end{figure*}

Figure \ref{fig:summary_profiles} shows the time-averaged density, temperature, CR pressure, wind speed, heating-to-cooling ratio, and effective CR equation of state $p_c(\rho)$ for each of the varied $\vesc$ runs (simulations 1 and 11-14 of Table \ref{tab:runs}); we will discuss the the time variability of the solutions in \S\ref{sec:timedep}. The simulations range from nearly isothermal at lower $\vesc$ to exhibiting a sharp temperature spike at higher $\vesc$, where the gas quickly heats up from $\sim 10^{4}-10^6$\,K near the base of the wind. At the fiducial $\vesc = 420$\,km/s but with varied base densities, CR pressures, and Alfv\'en speeds, the profiles are similar to the fiducial result, though the spike in temperature occurs at a slightly different location (see Figure \ref{fig:T_spike_approx}), and the outflow speed and mass loss rates of the outflow vary modestly (see Table \ref{tab:runs}).

Overall, we find that the depth of the gravitational potential is the most significant parameter in determining the outflow properties. Physically, this is because the stronger gravity solutions (higher $\vesc$) have much smaller gas density scale heights and thus much lower densities just exterior to the base of the wind. This leads to cooling being less important relative to heating of the gas by the streaming CRs. Once CR heating drives the gas temperature $\gtrsim 10^4$\,K, the putative balance between CR heating and radiative cooling becomes thermally unstable. The relative noisiness of the time-averaged $\vesc = 250$\,km/s, $\vesc=180$\,km/s, and $\vesc=130$\,km/s profiles is due to this instability occurring over an extended region, resulting in a much more time-variable solution, as we discuss in detail in \S\ref{sec:timedep}.

Steady-state CR wind theory predicts that $p_c \propto \rho^{2/3}$ when $v \ll v_A$ and $p_c \propto \rho^{4/3}$ when $v \gg v_A$ (equation \ref{eq:dpcdr}). Figure \ref{fig:summary_profiles} shows that $p_c \propto \rho^{2/3}$ is indeed satisfied at high densities near the base when $v \ll v_A$, though at larger radii (lower densities), a $p_c\propto\rho$ scaling is visible instead of $p_c \propto \rho^{4/3}$; this is because $v \sim v_A$. In addition, a $p_c\propto\rho^{1/2}$ scaling is apparent at intermediate radii where thermal instability begins to occur and the fluctuations in gas density are largest (see \S\ref{sec:timedep}). \citet{Quataert_streaming} showed that $p_c \propto \rho^{1/2}$ is a consequence of strong CR bottlenecks; see \cite{Tsung_Oh_Jiang_Staircase} for related arguments. They further argued that the larger CR pressure implied by $p_c \propto \rho^{1/2}$ (in comparison to $p_c \propto \rho^{2/3}$) leads to stronger galactic winds than predicted by standard CR wind theory. In \citet{Quataert_streaming}'s simulations, however, $p_c \propto \rho^{1/2}$ was present over a larger range of radii than we find here. This is a consequence of the more realistic thermodynamics in the present simulations; the instabilities leading to CR bottlenecks are suppressed once gas pressure becomes dynamically more important exterior to the temperature spikes in Figure \ref{fig:summary_profiles} (see \S \ref{sec:timedep} for more discussion).

\begin{figure}
    \centering
    \includegraphics[width=0.48\textwidth]{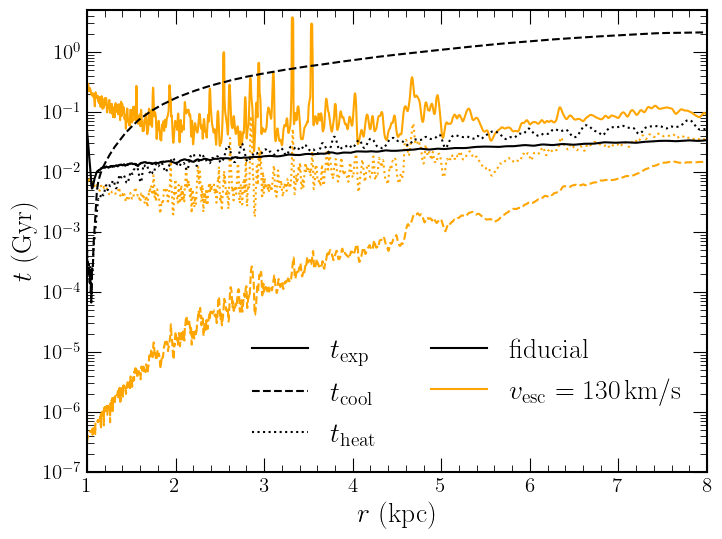}
    \caption{The expansion (solid), heating (dotted), and cooling (dashed) timescales as functions of radius for the fiducial $\vesc = 420\,$km/s (black) simulation and the $\vesc=130$\,km/s (orange) simulation. In the fiducial case, the cooling time is initially the shortest but quickly the solution transitions to one in which heating overwhelms cooling, leading to the strong spike in temperature seen in Figure \ref{fig:summary_profiles}. For the $\vesc = 130$\,km/s model, the cooling timescale is always significantly shorter than the heating and expansion timescales and the solution is roughly isothermal.}
    \label{fig:energy_timescales}
\end{figure}

Figure \ref{fig:summary_profiles} demonstrates that as the gas moves further out of the galaxy, the thermally unstable solutions are inevitably driven to lower densities and higher temperatures at which cooling is less dynamically important. In reality, the thermally unstable solutions that we find here are likely to have a rich multiphase structure not captured in our 1D simulations; we return to this in \S\ref{sec:discussion}. To more quantitatively describe the thermodynamics of the outflow and the importance of cooling, Figure \ref{fig:energy_timescales} compares the expansion timescale $t_{\mathrm{exp}} \equiv H/v$ (where $H$ is the density scale height $(\md\log\rho/\md r)^{-1}$), the cooling timescale $t_\mathrm{cool} \equiv p/(n^2\Lambda)$, and the CR heating timescale of the gas $t_\mathrm{heat} \equiv p/|v_A\md p_c/\md r|$. Close to the base of the wind, $t_{\rm cool} \lesssim t_{\rm heat} \ll t_{\rm exp}$. This is the approximately isothermal region in which cooling and heating balance, as described analytically in \S\ref{sec:base}. For $\vesc=130$\,km/s this hierarchy of timescales is maintained throughout the flow and the solution remains nearly isothermal everywhere. Note, however, that $t_{\rm cool}/t_{\rm exp}$ still increases substantially with radius for $\vesc = 130$\,km/s, and eventually, at sufficiently large radii, the gas density would decrease to the point that cooling would become negligible. For winds with stronger gravity such as the fiducial $\vesc=420$\,km/s, however, the cooling time increases much more rapidly with increasing radius and cooling is negligible exterior to $\sim 1.5$\,kpc. Effectively, this exterior solution at larger radii is well modeled using \citet{Ipavich1975}'s original CR streaming wind solutions that entirely neglect radiative cooling. We also note that while the outward expansion time of the CRs in the fiducial $\vesc=420$\,km/s model in Figure \ref{fig:energy_timescales} is comparable to the pion loss time of $\approx 5\times 10^{-2} (n/\,{\rm cm^{-3}})^{-1}\,$Gyr at the base, because the density falls rapidly, the pion loss time sharply rises after $\sim 0.01\,$kpc and so neglect of pionic losses is self-consistent. This is only borderline true in the $\vesc = 130\,$km/s galaxy, which has a longer wind expansion time due to the shallower potential.

\begin{figure}
    \centering
    \includegraphics[width=0.48\textwidth]{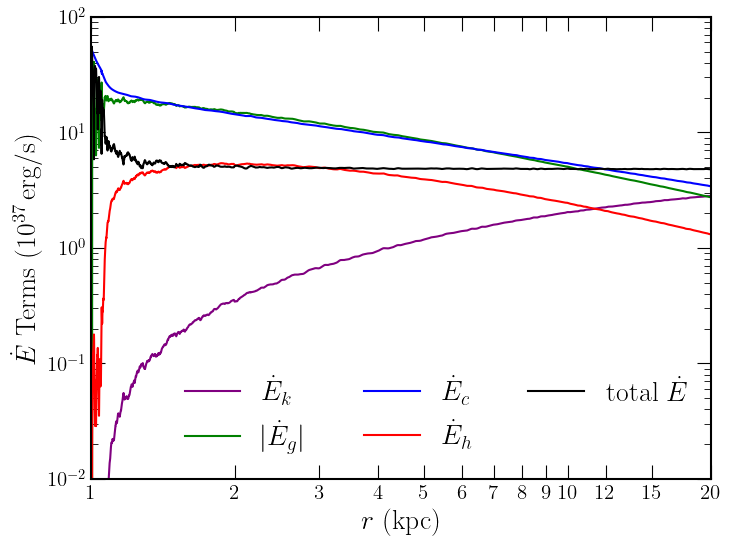}
    \caption{The gas kinetic energy flux (purple), gravitational energy flux (green), gas enthalpy flux (red), CR energy flux (blue), and total energy flux (black) as defined in equation \ref{eq:Edot} for the fiducial $\vesc = 420\,$km/s model in the larger $100\,$kpc box. The near equivalence of the gravitational and CR energy fluxes leads to a simple expression for the mass-outflow rate (see \S \ref{sec:mdot}). Note also that the asymptotic energy flux is only $\sim 10\%$ of the input energy flux, which corresponds to inefficient preventive feedback at larger radii.}
    \label{fig:Edot_terms}
\end{figure}

Figure \ref{fig:Edot_terms} shows the contributions of each term in equation \ref{eq:Edot} to the total energy flux of the wind in the fiducial model in the larger $100\,$kpc box. For reference, the initial input CR power is given by
\begin{align}
    \label{eq:Edotc0}
    \dot{E}_{c0} & = 16\pi r_0^2p_{c0}v_{A0} \nonumber \\
    & \approx 7.7\times 10^{38}\,\mathrm{erg/s}\left(\frac{r_0}{1\,\mathrm{kpc}}\right)^{2}\left(\frac{p_{c0}}{1\,\mathrm{eV/cm}^3}\right)\left(\frac{v_{A0}}{10\,\mathrm{km/s}}\right)
\end{align}
The total $\dot{E}$ decreases sharply at small radii near the base of the wind due to cooling, but remains constant for $r\gtrsim 2$\,kpc once the density has decreased significantly (so the effect of cooling becomes negligible). When the wind speed is low, the dominant contributions to $\dot{E}$ are the CR energy flux and the gravitational energy flux, which balance each other across a wide range of radii. In \S\ref{sec:mdot} we will use this feature of the wind to estimate its mass loss rate. The gas enthalpy flux rises rapidly following the onset of thermal instability due to the sharp spike in temperature and gas pressure at small radii. At larger radii, the wind accelerates primarily due to gas pressure while the temperature falls; the kinetic energy flux eventually dominates over the gas enthalpy flux.

The asymptotic wind power for the fiducial model shown in Figure \ref{fig:Edot_terms} is only $\sim 10\%$ of the input CR power at the base of the wind. Some of the input energy is lost radiatively, but most is lost to gravity driving the wind to large radii; this is reflected in $\dot E_c \simeq \dot E_g$ in Figure \ref{fig:Edot_terms}.  We find a similar ratio of the asymptotic wind power to the input CR power in all of our simulations (see the last column of Table \ref{tab:runs}). Since the energy per supernovae (SNe) supplied to CRs is $\sim 10\%$, this implies that the asymptotic wind power found here is only $\sim 1\%$ of the SNe power. This is unlikely to have a significant dynamical impact on the surrounding circumgalactic medium (CGM), i.e., the ``preventive'' feedback due to the winds found here will be minor.

\begin{figure}
    \centering
    \includegraphics[width=0.48\textwidth]{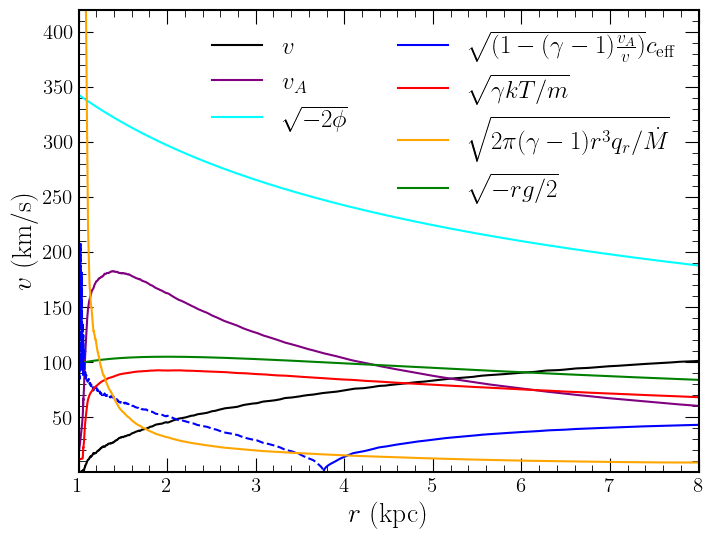}
    \caption{The wind speed (black), Alfv\'en speed (purple), local escape velocity (cyan), and components of the numerator and denominator critical point speeds $v_d^2$ and $v_n^2$ (equations \ref{eq:vd} and \ref{eq:vn}) for the fiducial $\vesc = 420\,$km/s model in the larger $100\,$kpc box. The blue curve shows the effective CR sound speed contribution, and the dashed part of the curve indicates when its square is negative, so it is imaginary. The red curve shows the gas sound speed contribution. The orange curve represents the contribution of cooling, and the green curve shows the gravitational velocity that appears in the numerator of the steady-state wind equation \ref{eq:v_profile}. Note the very slow acceleration of the wind and that the solutions are magnetically dominated out to large radii.}
    \label{fig:velocities}
\end{figure}

Figure \ref{fig:velocities} illustrates the acceleration of the wind for the fiducial model in the larger $100$\,kpc box, in comparison to the Alfv\'en speed and the local escape speed, with the individual components of the numerator $v_n^2$ and denominator $v_d^2$ in the wind equation \ref{eq:v_profile} also plotted for reference. Note that the local escape speed $\sqrt{-2\phi(r)}$ is distinct from the $\vesc$ parameter used in our parametrization of the Hernquist potential; the latter is the escape speed from $r = 0$. From the base to $r\approx 2$\,kpc, we see that $v \ll v_A$, so we are well-justified in utilizing that approximation throughout \S\ref{sec:analytics}. As anticipated in \S\ref{sec:cooling}, closest to the base, when $v_d^2 < 0$ (i.e. the blue dashed curve is greater than the red curve), $v_n^2 < 0$ as well (the orange curve is greater than the green curve). Although $v_d^2$ becomes positive at $r\approx 1.1$\,kpc, $v_n^2$ remains negative until $r \approx 1.2$\,kpc. Because this occurs when $|v_d^2| < v^2$, though, per equation \ref{eq:v_profile}, the wind is able to continue accelerating. This demonstrates how the time-averaged numerical solution manages to evade the conceptual difficulties highlighted in \S\ref{sec:cooling} and accelerate outwards continuously despite having $v \ll v_A$ at its base. Another notable feature of Figure \ref{fig:velocities} is that the solution is magnetically dominated ($v_A \gtrsim c_s$) out to $\sim 5\,$kpc.

Ultimately, the speed in the fiducial wind exceeds the local escape velocity by $r\approx 20$\,kpc, and continues to accelerate before reaching a final velocity of approximately $150$\,km/s near the boundary of our larger domain. We note that in each of the models studied in the larger simulation box of $100\,$kpc, the wind continues to accelerate beyond the speeds achieved by the same model in the $10\,$kpc box. To characterize the maximum speed achievable by the wind for models in which we do not simulate a larger box, Table \ref{tab:runs} also includes a value of
\begin{equation}
    \label{eq:v_infty}
    v_\infty \equiv \left(v^2 + \frac{2\gamma}{\gamma-1}\frac{p}{\rho}\right)^{1/2}
\end{equation}
for each wind, incorporating the enthalpy contributions to the wind's specific energy. As a check, the value calculated for $v_\infty$ in the larger boxes closely matches the value calculated for the smaller boxes, as well as the velocity of the wind near the outer boundary in the larger boxes. As the wind accelerates to approach $v_\infty$, it passes through two critical points, at $r \approx 1.15$\,kpc and $r \approx 6$\,kpc for the fiducial model, where the speed is equal to the local values of $v_n$ or $v_d$. These critical points may be read off of Figure \ref{fig:velocities} as the locations at which the square of the black curve is equal to the sum of the square of the blue and red curves ($v^2 = v_d^2$) or the difference of the squares of the green and orange curves ($v^2 = v_n^2$). We discuss the critical points observed in each simulation in further detail in \S\ref{sec:critical_sim}.

\subsection{Time Dependence}
\label{sec:timedep}

\begin{figure*}
    \centering
    \includegraphics[width=0.95\textwidth]{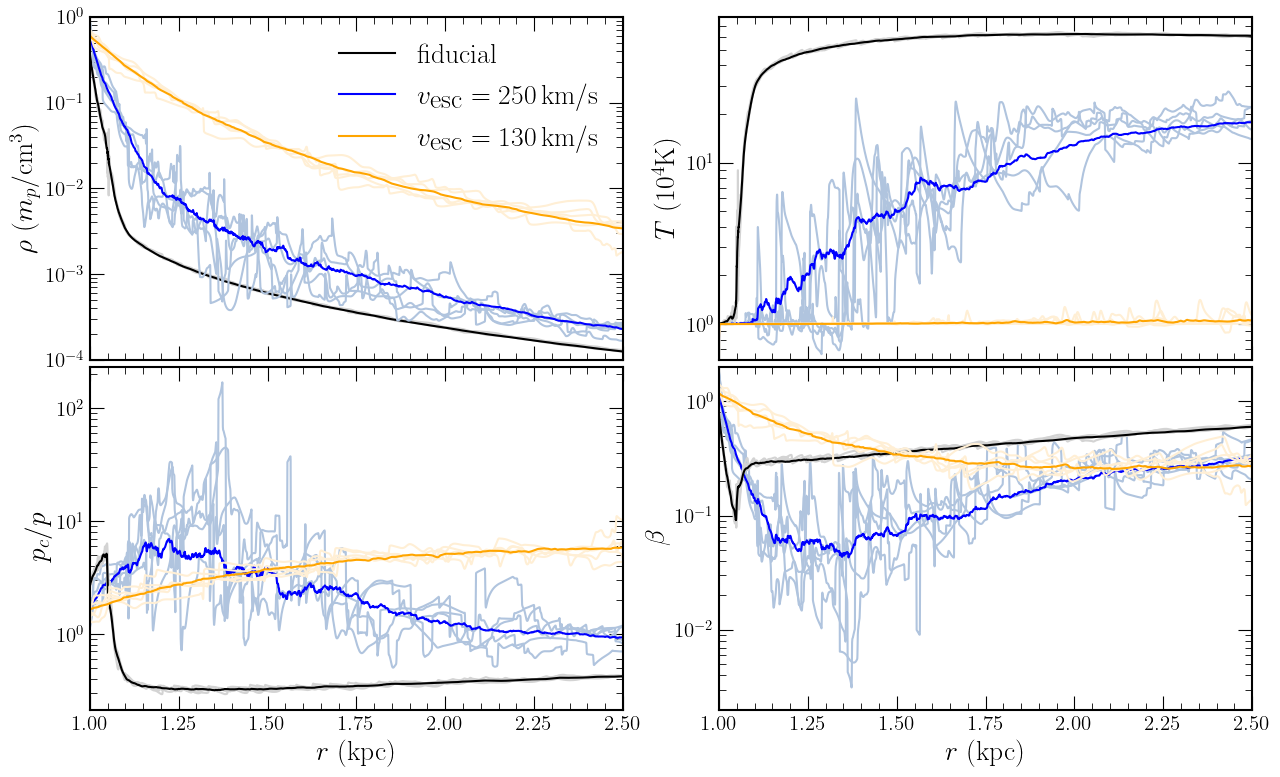}
    \caption{Profiles of the fiducial $\vesc = 420\,$km/s (black), $\vesc=250$\,km/s (blue), and $\vesc=130$\,km/s (orange) simulation runs demonstrating the time dependence of the wind's density (upper left), temperature (upper right), CR-to-gas pressure ratio (lower left), and gas-to-magnetic pressure ratio $\beta \equiv 2a^2/v_A^2$ (lower right) near the base. The lighter colored profiles are five different sample snapshots in which the wind is in statistical steady-state, and the darker colored profile is the time average. The individual snapshots show much more variation in the $\vesc=250$\,km/s run, and the time-averaged profile is correspondingly noisier. This large amplitude variability is due to thermal instability. For the $\vesc=130$\,km/s and $\vesc=420$\,km/s runs, much of the variability is likely due to acoustic instabilities, though this is much smaller in amplitude than the variability produced by thermal instability.}
    \label{fig:time_dependence}
\end{figure*}

In the numerical simulations, as $\vesc$ is reduced, a transition occurs between the smooth profiles of the fiducial or $\vesc = 330$\,km/s models and the much more variable  $\vesc = 250$ and $180$\,km/s models; this is evident even in the time-averaged profiles in Figure \ref{fig:summary_profiles}. The origin of this difference is the larger amplitude time variability that is introduced by the thermal instability in the lower $\vesc$ models. Figure \ref{fig:time_dependence} highlights this strong time dependence by comparing the time-averaged density, temperature, CR pressure to gas pressure ratio, and plasma $\beta = 2a^2/v_A^2$ to individual time snapshots. Although the snapshots were chosen at times where the profiles are observed to be in statistical steady-state (i.e. averages over randomly chosen time intervals have the same pointwise statistics), in the $\vesc = 250$\,km/s model especially, there is significant time variability.

The $\vesc = 250$\,km/s model represents an intermediate regime to the fiducial and $\vesc = 130$\,km/s models. In the fiducial model, thermal instability sets in quickly because cooling is only important for a small range of radii near the base, while for $\vesc = 130$\,km/s, cooling is important across the entire simulation domain and thermal instability largely does not set in, since the gas remains on the thermally stable part of the cooling curve at $T \approx 10^4$\,K. Figure \ref{fig:time_dependence} shows that the more gradually rising time-averaged temperature profiles with $\vesc \leq 250$\,km/s can now be identified as an average over a number of thermal-instability-induced temperature spikes that are similar to those in the fiducial and $\vesc = 330$\,km/s models, just across a larger range of radii. Eventually, at large radii, cooling gradually becomes less important even for the $\vesc=250\,$ km/s solution, and the wind solution is significantly more stable at a higher temperature (as observed in the time averages in Figure \ref{fig:summary_profiles}). The large variations in the gas density seen in Figure \ref{fig:time_dependence} also lead to corresponding fluctuations in the CR pressure, analogous to the ``staircase'' structure observed in \citealt{Tsung_Oh_Jiang_Staircase}. In the bottom left panel of Figure \ref{fig:time_dependence}, we see that these fluctuations lead the ratio of CR and gas pressure to vary across nearly two orders of magnitude in just a $1\,$kpc region near the base for the $\vesc = 250$\,km/s model, while $p_c/p\sim \mathcal{O}(1)$ in the other solutions.

\begin{figure}
    \centering
    \includegraphics[width=0.48\textwidth]{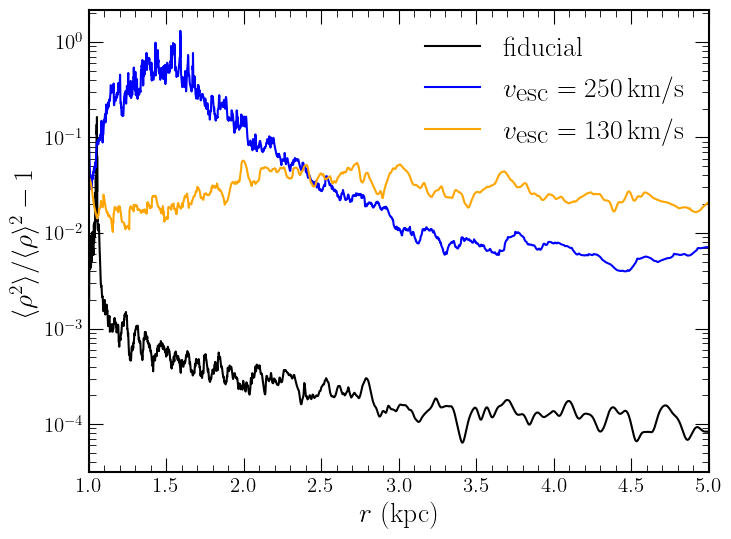}
    \caption{The normalized temporal variance in density for the fiducial $\vesc = 420\,$km/s (black), $\vesc=250$\,km/s (blue), and $\vesc=130$\,km/s (orange) models. In the fiducial model, the variance rises sharply at the location of the temperature spike, which demonstrates the onset of the thermal instability. In the $\vesc = 250$\,km/s case, the rise in variance is much broader, because the thermal instability is not as localized in radius as in the $\vesc=420$\,km/s model. Thermal instability is not present in the $\vesc=130$\,km/s model, for which the variability is likely due to sound wave instabilities.}
    \label{fig:rho_variance}
\end{figure}

Next, we quantify the radii at which the thermal instability has the greatest influence on the wind. Figure \ref{fig:rho_variance} shows the pointwise temporal variance in the density normalized by the time-averaged profile for the fiducial, $\vesc = 130$\,km/s, and $\vesc = 250$\,km/s simulations. While the fiducial model exhibits significant variation in time only where the temperature increases dramatically (visible as a spike in the curve at $r\approx 1.05$\,kpc), the higher variance in the $\vesc=250$\,km/s model extends throughout the simulation domain, and is substantial through $r\approx 3$\,kpc. In the $\vesc = 130$\,km/s model, while the variance is higher in general, there is no clear rise in the variance: because cooling is always more relevant, thermal instability does not set in. The density fluctuations are rarely spatially extended, typically occurring over scales of just $\sim 10$\,pc.

Although we identify thermal instability as the source of largest time variability in our models, several other instabilities are also likely to be present. In particular, the linear acoustic instabilities studied in \cite{Quataert_streaming} (in isothermal winds, when background gradients are present) and in \cite{Begelman1994} (in plasmas with CR-streaming-mediated heating, but without any cooling) are both realizable. The former occurs in the approximately isothermal region nearest the base, and is likely responsible for the small initial fluctuations in temperature that result in the onset of thermal instability. The change in the time-averaged CR equation of state from $p_c\propto\rho^{2/3}$ to $p_c\propto\rho^{1/2}$ observed in \cite{Quataert_streaming} as a consequence of CR bottlenecks is present in our models as well (see the lower right panel of Figure \ref{fig:summary_profiles}). However, it is driven more by thermal instability here than by the acoustic instabilities, in part because the latter are suppressed if gas pressure dominates.

We expect the \cite{Begelman1994} instability to potentially occur at large radii away from the base where cooling is less important. All winds studied here include regions in which $\beta \lesssim 0.2$ and $\beta \lesssim 0.5$, so that both forward and backward propagating acoustic waves may be unstable \citep{Begelman1994}. These instabilities are likely responsible for the small variations visible in e.g. the fiducial model in the $p_c/p$ panel beyond the temperature spike. However, we emphasize that for the parameter space considered in our models, thermal instability leads to much more dramatic variability, and is likely to have many more important observational and dynamical consequences than the acoustic wave instabilities. For instance, Figure \ref{fig:rho_variance} shows that although there is a baseline variance in each of the models over a wide range of radii (likely due to acoustic instabilities), the variance due to the thermal instability far exceeds that produced by the acoustic instabilities.

\subsection{Critical Points}
\label{sec:critical_sim}

\begin{figure*}
    \centering
    \includegraphics[width=0.95\textwidth]{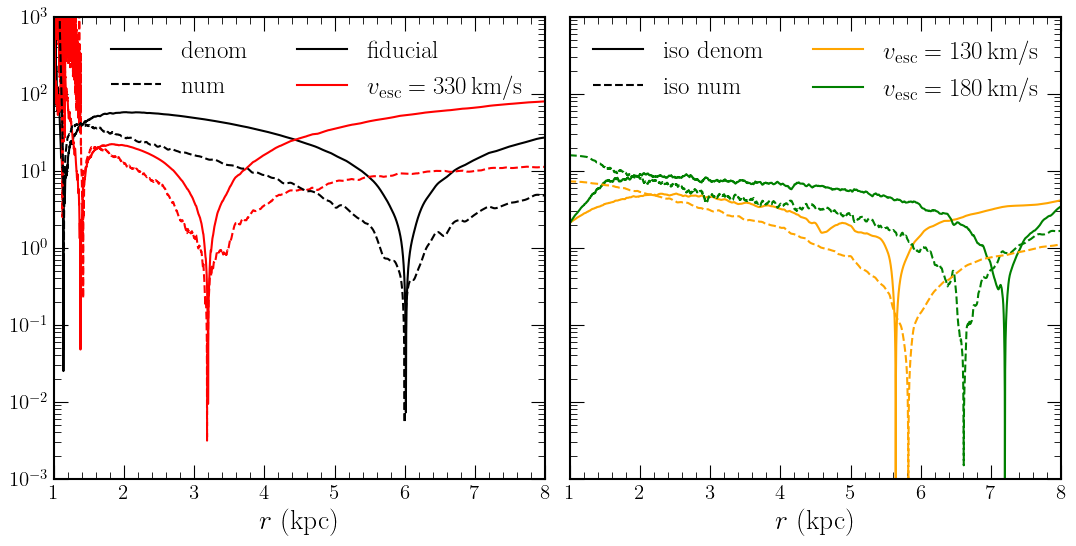}
    \caption{The numerator and denominator of the wind equation for the $\vesc = 130$ (orange), 180 (green), 330 (red), and fiducial 420\,km/s (black) models. The left panel shows the numerator ($v_d^2 - v_n^2$, dashed) and denominator ($v^2 - v_d^2$, solid) expressions calculated using equations \ref{eq:vd} and \ref{eq:vn}. Two critical points are present as the wind speed passes through the roots $v_{c\,+}$ and then $v_{c\,-}$ as expected from equation \ref{eq:crit_vel}. The right panel shows the numerator (dashed) and denominator (solid) expressions as calculated from the isothermal limit of equation \ref{eq:v_profile} instead; only one critical point is present in this limit.}
    \label{fig:crit_pt_splitting}
\end{figure*}

We find that the time-averaged profiles of our numerical solutions do pass through the analytically expected critical point(s), even though the flow is only steady in a statistical sense. Figure \ref{fig:crit_pt_splitting} shows the wind equation numerator and denominator expressions, $v_d^2 - v_n^2$ and $v^2 - v_d^2$ respectively from equation \ref{eq:v_profile}. Critical points are apparent as sharp dips in the plot when both the numerator and denominator expressions are simultaneously near-zero. The left panel demonstrates the presence of 2 critical points for the higher $\vesc$ models, while in the lower $\vesc$ models in the right panel, only 1 critical point is present. The latter is because the low $\vesc$ solutions correspond to the nearly isothermal limit in which there is indeed only one critical point (see \S\ref{sec:iso}). Although the full numerator and denominator expressions apply to the $\vesc = 130$ and 180\,km/s models, for those nearly isothermal solutions, we directly plot the numerator and denominator of the isothermal limit of the wind equation for clarity, because fluctuations due to cooling add significant noise to the numerator expression, making the zero harder to distinguish. As a check of the validity of this analysis method, we note that attempting to use the isothermal expressions for the higher $\vesc$ models in the left panel does not yield aligned zeros of the numerator and denominator, demonstrating that the full critical point expressions are important for those critical points. We also do not plot the intermediate-regime $\vesc = 250$\,km/s model in either panel, because it does not exhibit clear isothermal critical points, and its time-averaged profile is too noisy to allow for clear identification of the full critical points due to the increased variability.

In our simulations exhibiting two critical points, the first occurs when $v = v_{c,\,-}$ and the second when $v = v_{c,\,+}$, where $v_{c,\,\pm}$ are the two roots of equation \ref{eq:crit_vel}. The two critical points in the $\vesc=420$ and $330$\,km/s models in Figure \ref{fig:crit_pt_splitting} are consistent with the surprising analytic critical point structure discussed in \S\ref{sec:critical}. The existence of two critical points is a consequence of the presence of both cooling and an imaginary CR sound speed at some radii. The inner $v = v_{c,\,-}$ critical point occurs when cooling is energetically important, and the position may be determined approximately by when $v_d^2$ and $v_n^2$ first rise above 0. The outer critical point is the more familiar Parker-type critical point in which $v = \sqrt{-rg/2}$, as observed in the fiducial model velocities in Figure \ref{fig:velocities}.

In the lower $\vesc$ solutions with just one critical point, the critical speeds are similar such that $v = v_{c,\,-} \approx v_{c,\,+}$. The single critical point in these cases corresponds to taking the isothermal limit of equation \ref{eq:crit_vel} as described in \S\ref{sec:iso}. The slightly larger deviations between the location of zeros of the numerator and denominator in the $\vesc=130$\,km/s and $\vesc=180$\,km/s models are due to the increased time variability in those cases compared to the highest $\vesc$ simulations shown in the left panel, and also because the winds are not exactly isothermal.

\section{Discussion}
\label{sec:discussion}

Informed by our numerical results, in the following sections we describe how to approximate the total mass loss rate of the wind as well as its maximum temperature and speed. We then discuss some of the possible observational signatures of the CR-driven winds found here and summarize limitations of our modeling, as well as possible areas for future work.

\subsection{Approximating Mass-Loss Rates and Outflow Speeds}
\label{sec:mdot}

The simulations in \S\ref{sec:numerical_sims} show that the asymptotic speeds of CR-driven winds from the warm ISM are relatively small compared to the initial escape speed. In this limit, the mass-loss rate is close to the maximum mass-loss rate allowed by energy conservation. That maximum rate is set by when the available energy primarily goes into lifting matter out of the gravitational potential, so that $|\dot E_g| \simeq \dot E_c$, i.e., $\dot M |\phi| \simeq \dot E_c$, as demonstrated explicitly in Figure \ref{fig:Edot_terms}. To estimate the resulting mass-loss rate, however, we have to account for the fact that cooling removes energy from the wind at small radii so that only a fraction of the initial energy in CRs at the base of the wind is available to drive gas to large radii. To do so, we will estimate the radius $r^*$ at which cooling becomes negligible. Such a radius does not exist for our lowest $v_{\rm esc} = 130$\,km/s simulation, which is nearly isothermal out to large radii. More appropriate analytic approximations for the mass-loss rate in this isothermal limit were given in \cite{Mao_Ostriker} and \cite{Quataert_streaming}.

Given an estimate of the radius $r^*$ at which cooling becomes subdominant, the net mass loss rate is roughly
\begin{align}
    \label{eq:Mdot_max_general}
    \dot{M} & \approx \dot{E}_c(r^*)/|\phi(r^*)| \\
    & = 16\pi (r^*)^2 p_c(r^*)v_A(r^*)/|\phi(r^*)|
\end{align}
Because in most solutions cooling is only important at relatively small radii, where $v \lesssim v_A$, we approximate $p_c\sim\rho^{2/3}$. Then, using the split-monopole configuration and Hernquist potential, we find
\begin{equation}
    \label{eq:Mdot_max_specific}
    \dot{M} \approx \dot{M}_\mathrm{ref} \left(1 + \frac{r^*}{b}\right)\left(\frac{\rho(r^*)}{\rho_0}\right)^{1/6}
\end{equation}
where we have defined a reference mass loss rate value,
\begin{align}
    \label{eq:Mdotref}
    \dot{M}_\mathrm{ref} & \equiv 32\pi \frac{r_0^2v_{A0}p_{c0}}{\vesc^2} \nonumber \\
    & \approx 0.04\,\frac{\Msun}{\mathrm{yr}}\left(\frac{r_0}{1\, \mathrm{kpc}}\right)^2\left(\frac{v_{A0}}{10\,{\rm km/s}}\right)\left(\frac{p_{c0}}{1 {\rm eV/cm^{3}}}\right)\left(\frac{\vesc}{250\,{\rm km/s}}\right)^{-2}
\end{align}
Note that while our models treat the base CR pressure and escape velocity as independent parameters, galaxies with larger escape velocities are likely to host increased star formation and thus maintain an increased base CR pressure as well. From equation \ref{eq:Mdotref}, therefore, in real winds we may expect the mass loss rate to have only a sub-linear scaling with $p_{c0}$ (or equivalently, a steeper than $\vesc^{-2}$ scaling with escape velocity).

To estimate the density in equation \ref{eq:Mdot_max_specific}, we use the analytic implicit isothermal solution from equation \ref{eq:implicit_rho}. Although this somewhat underestimates the true density in the simulations (see Figure \ref{fig:T_spike_approx}), the weak $\propto \rho(r^*)^{1/6}$ scaling in equation \ref{eq:Mdot_max_specific} implies that our estimate of $\dot M$ is not that sensitive to this uncertainty in the density profile.

It then only remains to estimate $r^*$. Because cooling enters the wind equation only through $v_n^2$, we estimate $r^*$ by estimating when $v_n^2(r^*) = 0$. For radii smaller than $r^*$, $v_n^2 < 0$ and $v_d^2 < 0$, but beyond $r^*$, we have solutions with $v_n^2 > 0$ matching more conventional wind expectations as discussed in \S\ref{sec:cooling}. Setting $v_n^2=0$ using equation \ref{eq:vn} then yields
\begin{align}
    \dot{M} & = \frac{4\pi(\gamma-1)(r^*)^2q_r(r^*)}{-g(r^*)} \\
    \label{eq:Mdot_vn0}
    & \simeq 4\pi (r^*)^2\rho(r^*) v_A(r^*)\frac{(\gamma-1)\ceffsq(r^*)}{\ceffsq(r^*) + a(r^*)^2}
\end{align}
where in the second line, we have used the heating-cooling balance and assumed hydrostatic equilibrium. Once again taking $p_c\propto\rho^{2/3}$ and assuming the split-monopole configuration, we ultimately arrive at the approximate mass loss rate
\begin{align}
    \label{eq:Mdot_vn_specific}
    \dot{M} \simeq \frac{8\pi}{3}(\gamma-1) \frac{r_0^2v_{A0}p_{c0}}{a_0^2} & \left(\frac{\rho(r^*)}{\rho_0}\right)^{1/6} \nonumber \\
    & \times \left(\frac{T(r^*)}{T_0} + \frac{2}{3}\frac{p_{c0}}{p_0}\left(\frac{\rho(r^*)}{\rho_0}\right)^{-1/3}\right)^{-1}
\end{align}
where $p_0 \equiv \rho_0 a_0^2$ is the base gas pressure. Comparing equations \ref{eq:Mdot_max_specific} and \ref{eq:Mdot_vn_specific}, we see that $r^*$ satisfies
\begin{equation}
    \label{eq:r_star}
    \left(1 + \frac{r^*}{b}\right)\left(\frac{T(r^*)}{T_0} + \frac{2}{3}\frac{p_{c0}}{p_0}\left(\frac{\rho(r^*)}{\rho_0}\right)^{-1/3}\right) = \frac{\gamma-1}{12}\frac{\vesc^2}{a_0^2}
\end{equation}
Equation \ref{eq:r_star} can be solved implicitly for $r^*$ given a temperature and density profile. We use the analytic approximations for $\rho(r)$ and $T(r)$ near the base of the wind from \S \ref{sec:base} to solve for $r^*$ and then $\dot M$.

\begin{figure}
    \centering
    \includegraphics[width=0.48\textwidth]{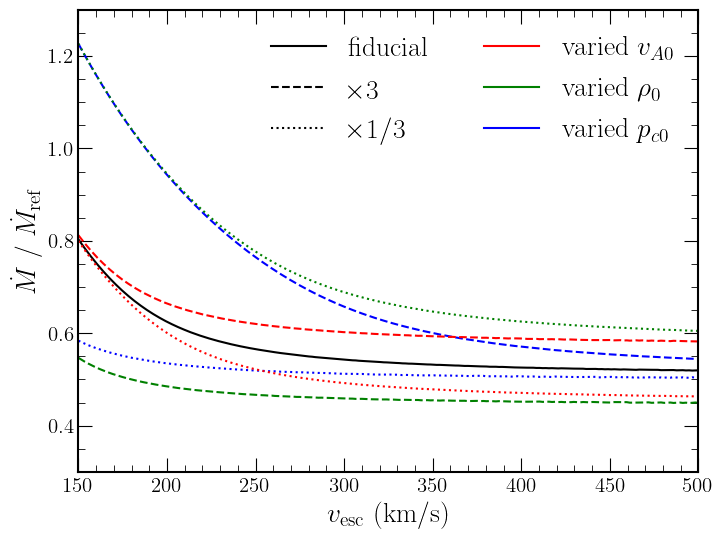}
    \caption{Analytic approximation to the mass-outflow rate $\dot{M}$ (equations \ref{eq:Mdot_max_specific} and \ref{eq:r_star}) as a function of $\vesc$, normalized by $\dot{M}_\mathrm{ref}$ (see equation \ref{eq:Mdotref}). The solid black curve is for the fiducial parameter choices; the dashed and dotted curves represent factor of 3 increases and decreases in $v_{A0}$ (red), $\rho_0$ (green), and $p_{c0}$ (blue) respectively.}
    \label{fig:Mdot_vesc_approx}
\end{figure}

\begin{figure}
    \centering
    \includegraphics[width = 0.48\textwidth]{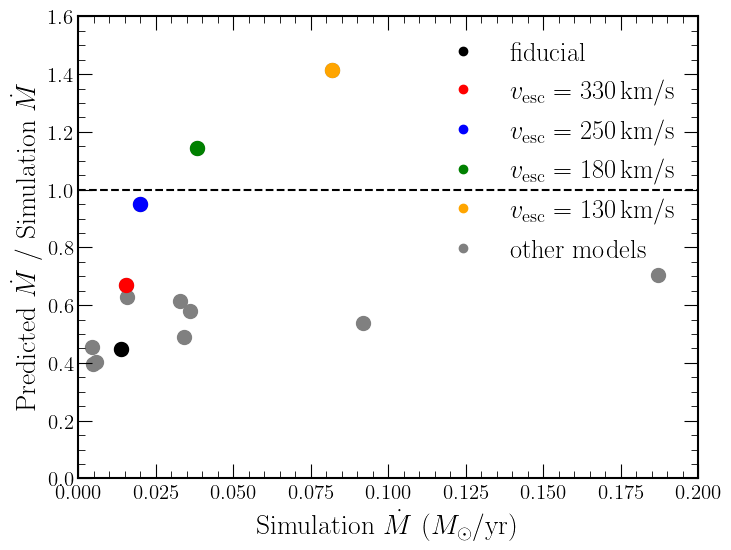}
    \caption{A comparison between the predicted (eqs. \ref{eq:Mdot_max_general} and \ref{eq:r_star}) and simulated mass loss rates. The black point is the fiducial $\vesc = 420\,$km/s simulation, the colored points indicate the varied $\vesc = 330$ (red), 250 (blue), 180 (green), and 130\,km/s (orange) simulations, and the grey points are  the remaining models.  The analytic estimate of the mass loss rate is good to about a factor of two.   Together with Figure \ref{fig:Mdot_vesc_approx}, this implies that equations \ref{eq:Mdotref} \& \ref{eq:MdrefvsMdstar} are a reasonable approximation of the mass loss rate in our CR-driven winds.}
    \label{fig:Mdot_comparison}
\end{figure}

Figure \ref{fig:Mdot_vesc_approx} shows the resulting predicted $\dot{M}$ as a function of $\vesc$ for several varied choices of base parameters, normalized by the ($\vesc$-dependent) reference value $\dot{M}_{\mathrm{ref}}$ from equation \ref{eq:Mdotref}. The dependence of $r^*$ on $\vesc$ alters the profiles from  $\dot{M}\propto\vesc^{-2}$ at low $\vesc$, but at high $\vesc$, cooling becomes unimportant almost immediately beyond the base, and so the scaling $\dot{M}\propto\vesc^{-2}$ becomes reasonably accurate. For the same reason, at higher $\vesc$, the dominant variation with all parameters is just through the prefactor term. Figure \ref{fig:Mdot_comparison} compares the predicted mass loss rates to the values realized in our simulations. The predicted values are correct within a factor of $\sim 2$, with deviations primarily due to errors in estimating $r^*$. The predictions are more accurate for the smaller $\vesc$ simulations because those density profiles are better-fit by the isothermal approximation of equation \ref{eq:implicit_rho}.

A useful alternative expression for the reference mass-loss rate is $\dot M_{\rm ref} \simeq 2 \dot E_{c0}/v_{\rm esc}^2$ where $\dot E_{c0} = 16 \pi r_0^2 v_{A0} p_{c0}$ is the total CR power of the galaxy. The latter can also be expressed as $\dot E_{c0} = \epsilon_c \dot M_* c^2$ where $\dot{M}_*$ is the star formation rate and $\epsilon_c \equiv 10^{-6.3} \epsilon_{c,\,-6.3}$ is set by the fraction of SNe energy that goes into CRs: for $10^{51}$\,erg per SNe and 1 SNe per 100 $M_\odot$ of stars formed, $\epsilon_c = 10^{-6.3}$ if $10\%$ of the SNe energy goes into primary CRs. The reference mass-loss rate in equation \ref{eq:Mdotref} can thus be rewritten as
\begin{equation}
\frac{\dot M_{\rm ref}}{\dot M_*} \simeq \frac{2 \epsilon_c c^2}{\vesc^2} \simeq 1.4 \, \epsilon_{c,-6.3} \, \left(\frac{v_{\rm esc}}{250\,\mathrm{km}/\mathrm{s}}\right)^{-2}
\label{eq:MdrefvsMdstar}
\end{equation}

Figures \ref{fig:Mdot_vesc_approx} and \ref{fig:Mdot_comparison} show that our simulations produce mass-loss rates $\dot M \sim \mathcal{O}(\dot M_{\rm ref})$ particularly at higher $v_{\rm esc}$. The mass-loss rates are somewhat higher for lower $\vesc$ because the density scale-height is larger and thus $\rho(r^*)/\rho_0$ is larger (see equation \ref{eq:Mdot_max_specific}). Our numerical results and equation \ref{eq:MdrefvsMdstar} thus imply that CR-driven winds can generate a mass loss rate of order or larger than the star-formation rate, particularly in lower mass galaxies. This is on the low end of the mass-loss rates required to reconcile the galaxy stellar-mass mass function and the halo mass function in cosmological galaxy formation models \citep{Somverville2015}. One unusual feature of our models is that the mass-loss is dominated by the warm ISM, yet the gas often ends up hotter than its initial temperature at large radii.

\begin{figure}
    \centering
    \includegraphics[width=0.48\textwidth]{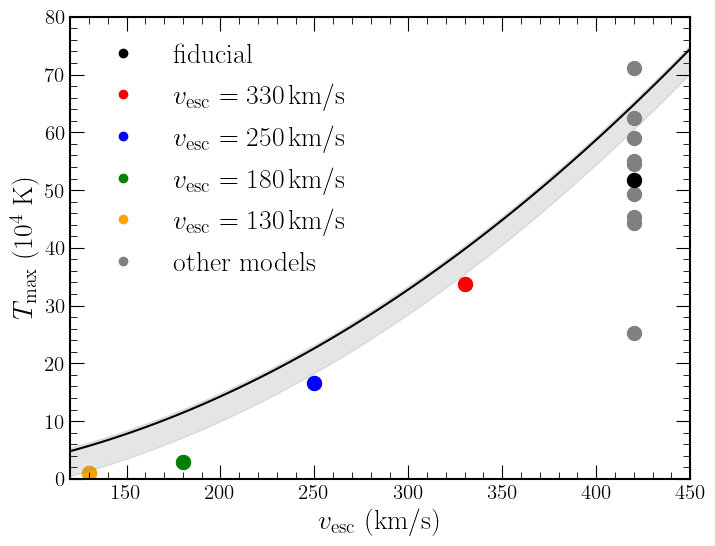}
    \caption{The maximum temperature predicted by equation \ref{eq:Tmax} compared to the maximum temperatures of the simulations. The black curve uses the fiducial values of $\rho_0$ and $p_{c0}$, while the grey shaded region indicates the upper bound predicted allowing for up to factor of 3 variations in $\rho_0$ and $p_{c0}$. The black point is the fiducial $\vesc = 420\,$km/s model, and the varied $\vesc = 330$ (red), 250 (blue), 180 (green) and 130\,km/s (orange) models are shown as colored points, while all other models are shown in grey. The one grey point exceeding the maximum estimate is the $r_0 = 5$\,kpc model (the last row in Table \ref{tab:runs}), and the grey point with the smallest maximum temperature is the $\rho_0 = 0.1\,$m$_p$/cm$^3$ model (the second-to-last row in Table \ref{tab:runs}).}
    \label{fig:Tmax_comparison}
\end{figure}

In addition to its value in calculating $\dot{M}$, equation \ref{eq:r_star} also allows for an estimate of the maximum temperature the wind will achieve, and hence its maximum speed. From the gas energy equation, the temperature profile is determined by the solution to
\begin{equation}
    \label{eq:dTdr}
    n v k \dv{T}{r} = (\gamma-1)(va^2-v_A\ceffsq)\dv{\rho}{r} - (\gamma-1)q_r
\end{equation}
and we see that the initial increase in temperature is driven by the fact that $va^2 \ll v_A\ceffsq$ near the base (since $\md\rho/\md r$ is always $< 0$). For radii beyond the predicted instability-driven temperature spike, however, we expect $va^2$ to become comparable to $v_A\ceffsq$ due to the increased temperature and accelerating wind speeds, and so the outer portion of the temperature profile must be decreasing. Therefore, we expect the largest temperatures to typically be reached at radii somewhat comparable to $r^*$, and we can use equation \ref{eq:r_star} to estimate a rough upper bound:
\begin{equation}
    \label{eq:Tmax}
    \frac{kT(r^*)}{m} < \frac{kT_\mathrm{max}}{m} \equiv  \left(\frac{\gamma-1}{12}\left(1 + \frac{r_0}{b}\right)^{-1}\vesc^2 - \frac{2}{3}\frac{p_{c0}}{\rho_0}\right)
\end{equation}
With $r_0/b$ held constant, we can identify an approximate $T_\mathrm{max}\propto\vesc^2$ scaling. We emphasize that equation \ref{eq:Tmax} assumes that $va^2 = v_A\ceffsq$ is satisfied near or interior to $r^*$; otherwise, the temperature could in principle continue to increase outwards. We do not have a rigorous proof that this ordering is satisfied but it is roughly true in our simulations: for reference, empirically, the simulations exhibit a $\sim 10\%$ increase in temperature beyond $T(r^*)$ before ultimately decreasing adiabatically at larger radii. However, the upper bound predicted by equation \ref{eq:Tmax} is still not saturated in most of our simulations. Figure \ref{fig:Tmax_comparison} compares the maximum temperature achieved by the simulated winds to the predicted $T_{\rm max}$ assuming fiducial parameter choices for $p_{c0}$ and $\rho_0$, as a function of $\vesc$. An approximate $T_\mathrm{max}\propto \vesc^2$ scaling is somewhat visible, although as shown by the wide range of maximum temperatures for the models with $\vesc = 420$\,km/s, variations in the base density, CR pressure, and Alfv\'en speed are also important in determining the maximum temperature the wind achieves.

\begin{figure}
    \centering
    \includegraphics[width=0.48\textwidth]{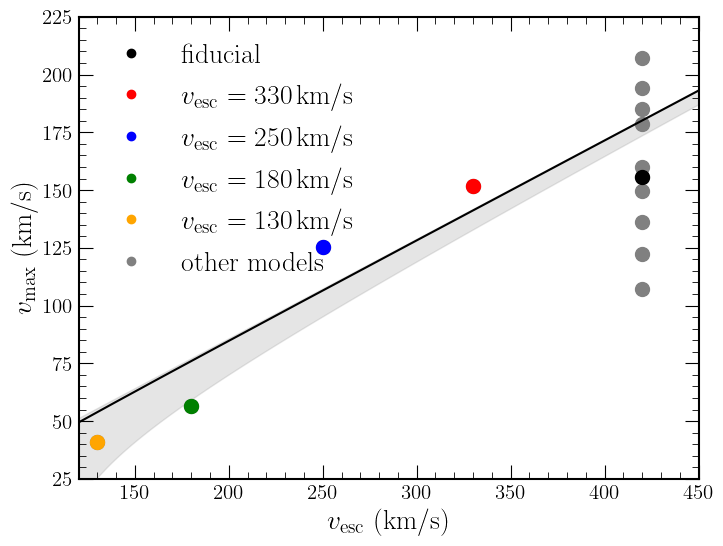}
    \caption{A comparison of the maximum wind speeds predicted using equation \ref{eq:vmax} and the simulated $v_\infty$ (from Table \ref{tab:runs}, calculated using equation \ref{eq:v_infty}) at $0.8r_\mathrm{out}$. Similar to Figure \ref{fig:Tmax_comparison}, the black curve uses the fiducial values of $\rho_0$ and $p_{c0}$, while the grey shaded region indicates the upper bound predicted allowing for up to factor of 3 variations in $\rho_0$ and $p_{c0}$. The black point is the fiducial $\vesc = 420\,$km/s simulation, the colored points indicate the varied $\vesc = 330$ (red), 250 (blue), 180 (green), and 130\,km/s (orange) simulations, and the grey points are all the remaining models.}
    \label{fig:vmax}
\end{figure}

Note that this upper bound predicts that the maximum sound speed of the gas is only a small fraction of $\vesc$. Under the assumption that by the time the temperature reaches its maximum, cooling is no longer energetically important, we may estimate the maximum speed achievable in the wind by assuming that the enthalpy of the wind is ultimately converted into kinetic energy:
\begin{align}
    \label{eq:vmax}
    v_\mathrm{max} & \simeq \left(\frac{2\gamma}{\gamma-1}\frac{kT_\mathrm{max}}{m}\right)^{1/2} \\
    & = \left(\frac{\gamma}{6}\left(1 + \frac{r_0}{b}\right)^{-1} - \frac{4\gamma}{3(\gamma-1)}\frac{p_{c0}}{\rho_0\vesc^2}\right)^{1/2}\vesc
\end{align}
To good approximation, because $\vesc^2 \gg p_{c0}/\rho_0$, we see that $v_\mathrm{max}$ scales linearly with $\vesc$, and for our models in which $\gamma = 5/3$ and $r_0/b = 1/2$, we expect $v_\mathrm{max}\approx 0.4\vesc$. Once again, although there is no rigorous proof that the maximum wind speed is determined by matching the enthalpy and kinetic energy fluxes of equation \ref{eq:Edot}, we find that this estimate is well-justified in our simulations because most of the winds become gas pressure dominated due to run-away CR heating as the density drops. The only exception to this is our nearly isothermal solutions at the lowest $v_{\rm esc}$ (for which the isothermal models of \citealt{Mao_Ostriker} and \citealt{Quataert_streaming} are a good analytic approximation). Figure \ref{fig:vmax} compares the predicted $v_{\rm max}$ from equation \ref{eq:vmax} to the simulated $v_\infty(r=0.8r_\mathrm{out})$ shown in Table \ref{tab:runs}. Overall, the estimated maximum wind speed matches the simulated value to within $\sim 50\%$, although there is significant variation among winds with the fiducial $\vesc = 420$\,km/s potential when other parameters are varied. The strong connection between the asymptotic wind speed and the galaxy escape speed found here is reminiscent of similar trends in observations (e.g. \citealt{Weiner2009}) although the normalization of our correlation between wind speed and escape speed is a factor of few lower than that observed.

\subsection{Emission and Absorption in the Wind}
\label{sec:observations}

The models presented in this paper are sufficiently idealized to preclude a detailed comparison to observations. Nonetheless, it is valuable to highlight a few features of the solutions found here that bear on observations of galactic winds with emission and absorption line diagnostics.

\begin{figure*}
    \centering
    \includegraphics[width=0.95\textwidth]{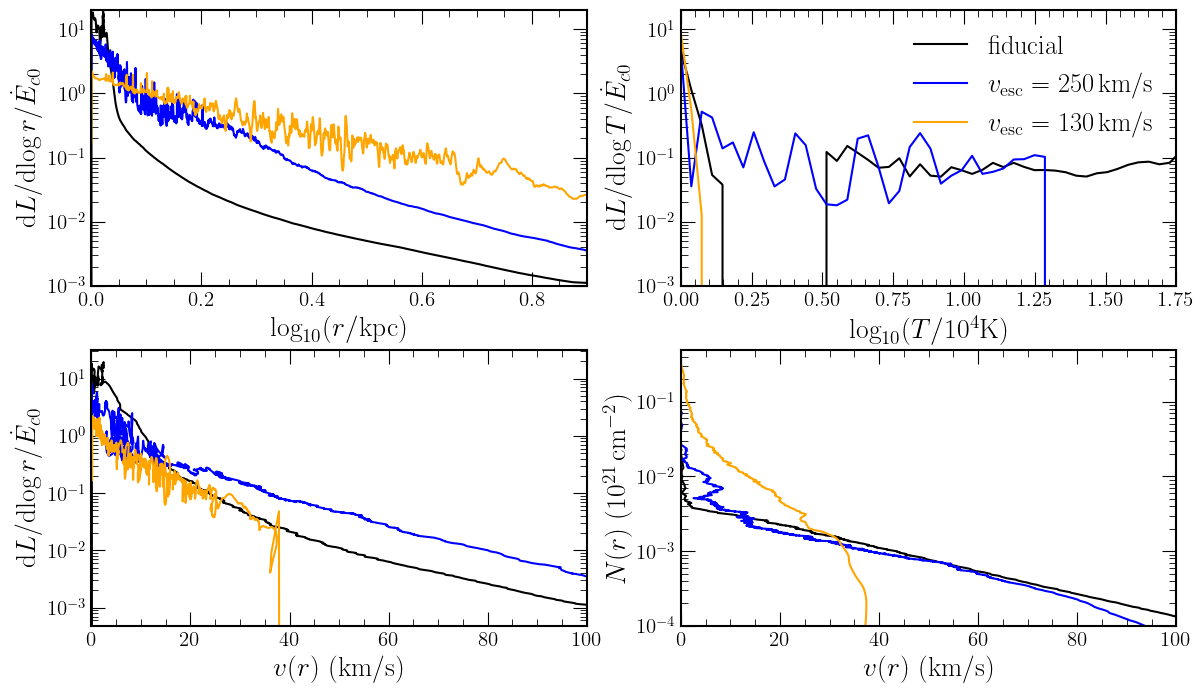}
    \caption{Proxy quantities for the emission as a function of radius (upper left), the emission as a function of temperature (upper right), emission as a function of wind speed (lower left), and absorption column density as a function of wind speed (lower right). In each panel, the black curve is the fiducial $\vesc = 420$\,km/s model, the blue curve is the $\vesc = 250\,$km/s model, and the orange curve is the $\vesc = 130$\,km/s model. For reference, in these models $\dot{E}_{c0} \approx 6.6\times10^{38}\,$erg/s (see equation \ref{eq:Edotc0}).}
    \label{fig:luminosity_plots}
\end{figure*}

To quantify the luminosity radiated by the outflowing wind, we define
\begin{equation}
    \label{eq:luminosity_r}
    L(r) \equiv \int_{r_0}^{r}4\pi r'^2\md r' q_r(\rho(r'), T(r'))
\end{equation}
and calculate the luminosity from logarithmically-spaced bins of radius and temperature, $\md L/\md\log r$ and $\md L/\md\log T$ respectively. These profiles normalized by the base CR power (see equation \ref{eq:Edotc0}) are shown for the fiducial $\vesc = 420$\,km/s, intermediate $\vesc = 250\,$km/s, and nearly isothermal $\vesc = 130$\,km/s models in the upper left and upper right panels of Figure \ref{fig:luminosity_plots} respectively. The upper left panel shows that emission from the gas is spatially extended, with $\sim 1-10$\% of the base CR power being emitted at radii of $\sim 5$ times the base radius of the wind. Note that this would correspond to outside the galaxy in any realistic galaxy model. This extended emission is a consequence of the interplay between CR heating and cooling of the wind and is thus a direct diagnostic of the physical origin of the wind. The upper right panel of Figure \ref{fig:luminosity_plots} shows that the emission in the nearly isothermal $\vesc = 130$\,km/s solution is dominated by the $\sim 10^4$\,K gas. By contrast, the other models show emission over a  wide range of temperatures from $10^4-$ a few $\times 10^5$\,K. Emission signatures would thus be present from the optical to the UV. Interestingly, however, the relatively low maximum temperatures highlighted in \S\ref{sec:mdot} imply that there would not be significant X-ray emission from these CR-driven winds. The lower left panel of Figure \ref{fig:luminosity_plots} shows $\md L/\md\log r$ as a function of radius plotted against the wind speed at that radius; the emission is velocity-resolved, and in the non-isothermal winds, a significant fraction is emitted at quite slow speeds of $\sim10\,$km/s.

To quantify the potential absorption signatures associated with the wind as viewed towards the central galaxy, we define the column density of the wind exterior to a given radius using
\begin{equation}
    \label{eq:column}
    N(r) \equiv \int_r^\infty \md r' n(r')
\end{equation}
The lower right panel of Figure \ref{fig:luminosity_plots} shows $N(r)$ as function of the wind speed at that radius $v(r)$ for the same fiducial, $\vesc = 250$\,km/s, and $\vesc = 130\,$km/s models shown in the left and middle panels. The wind speed here is a proxy for the range of Doppler shifted wavelengths that would be present in absorption line diagnostics of the wind. The column density profiles of the nearly isothermal wind is steeper and more sharply peaked at the base of the wind than those of the winds involving thermal instability, but among the non-isothermal winds, the columns are similar across a wide range of $\vesc$. We stress, however, that for resonance lines with high cross sections, the observed absorption line depth will depend strongly on the covering fraction as a function of velocity, not the column density we show in the right panel of Figure \ref{fig:luminosity_plots}. Of course, our 1D models cannot predict this covering fraction.

Overall, many of these features correspond well with the luminous, extended emission from diffuse ionized gas present in low-mass and starburst galaxies (e.g. \citealt{Heckman2015}, \citealt{McQuinn19}, \citealt{Marasco_ShakenBlown}, \citealt{Rautio_2022}, \citealt{Lu2023_eDIG}, \citealt{Xu2023}). In several of these galaxies, $\sim0.1\%$ of the bolometric luminosity can be observed as far as $6\,$kpc from the galaxy's center. The typical wind speed measured in such systems may also be as low as $10-100$\,km/s, consistent with our models.

Finally, we emphasize that the key feature of many of the solutions presented here is that thermal instability is important near the base of the wind. The nonlinear outcome of thermal instability is not well modeled in 1D but the outcome is undoubtedly that a multiphase medium develops for the radii in the wind where CR heating and radiative cooling are both important (see \citealt{HuangDavisThermal}). The resulting mulitphase wind solutions are likely to have many applications to understanding the complex phase structure seen in real galactic winds \citep{Veilleux2005}, but we defer more detailed calculations of the predicted emission and absorption line signatures to future multi-dimensional simulations.

\subsection{Caveats and Generalizations}
\label{sec:caveats_generalizations}

The goal of this work has been to explore the rather subtle interplay between CR-streaming-mediated heating and radiative cooling in galactic winds. For this purpose, the idealized spherically symmetric wind models presented here allow rapid exploration of parameter space and some analytic progress in modeling the wind solutions. The downside is that we have made a number of simplifications that limit the quantitative applicability of our results to realistic winds. 

First and foremost, we expect the gas to develop a complex 3D multiphase structure due to thermal instability, as in \cite{HuangDavisThermal}: the significant variance in density and temperature near the radii where the instability occurs (as seen in Figures \ref{fig:time_dependence} and \ref{fig:rho_variance}) is the manifestation of this in our 1D solutions. Additionally, we neglect thermal conduction, which may play an important role in the thermal structure of the wind due to the strong temperature gradients associated with CR heating overwhelming cooling (upper right panel of Figure \ref{fig:summary_profiles}). Thermal conduction is also important for setting a physical length-scale for the thermal instability. It is striking that none of our models with thermal instability produce co-spatial cold and hot gas at large radii: CR heating always eventually overwhelms cooling. This is different from other simulations of cold clouds embedded in a CR filled medium in which the bottleneck effect suppresses CR heating in the cold clouds and enables them to be accelerated intact (e.g. \citealt{Wiener2017,Wiener2019,HuangDavisThermal}). It is unclear if this difference is a result of the much larger dynamic range in density, temperature, and radii simulated here or a limitation of poorly resolved cold clouds in our simulations without thermal conduction. Our results are statistically converged at the resolutions simulated here, but this does not guarantee that inclusion of additional physics like conduction will lead to the same result or the same convergence properties.

Our simulations did not include CR diffusion in addition to CR streaming.  The motivation for this is that there are good observational and theoretical arguments that the bulk of the CR population (with energies $\sim$ GeV) can be self-confined by the streaming instability (e.g., \citealt{Blasi2012}). That said, there is also in general a correction to the pure CR streaming flux we have assumed here (the form of this correction is subtle and typically does not take the form of a diffusive flux; e.g., \citealt{Wiener2019, Kempski2022}). Future calculations of galactic winds with a full treatment of this streaming correction would be valuable. 

Another important aspect of multi-dimensional simulations is that the geometric expansion of the wind in a disk-like geometry is more gradual than in the spherical split-monopole setup considered in this paper. As a result, the gas density is likely to decrease more slowly away from the galaxy. This would move the transition where CR heating exceeds radiative cooling out in radius relative to the models presented here. We also note that the simulations found here are magnetically dominated over a large range of radii (see Figures \ref{fig:velocities} and \ref{fig:time_dependence}). As a result, it is likely that there is a combination of closed and open field lines that can only be modeled using more realistic multidimensional simulations.

One somewhat unusual property of the wind solutions found here is that the asymptotic wind speeds are quite slow: they remain factors of few below the escape speed from $r = 0$, as shown in Figure \ref{fig:vmax}. The acceleration of the wind in radius is also relatively gradual, as shown in Figure \ref{fig:summary_profiles}. For massive galaxies, the significant bounding pressure of the ambient CGM might therefore inhibit wind propagation into the CGM, particularly since our outflows are not supersonic with respect to the virialized gas in the CGM. Calculations with a more realistic ambient CGM would be valuable for determining its effects on the properties of CR-driven winds.

Finally, we note that in all of our models, we neglect additional sources of gas heating and cooling, including photo-heating from starlight, which is important in the warm ISM. \citet{Wiener_Zweibel_2013} show in models of the Milky Way that CR heating becomes increasingly important relative to photo-heating above the midplane of the disk. Our models are a natural extension of their solutions to even larger heights where CR heating dominates. It would, however, be valuable to carry out more complete calculations including both photo-heating and CR heating. Explicit inclusion of pionic losses in the CR energy equation would also be useful to include since the slow wind speeds found here imply that pionic losses can be important in star-forming galaxies with a dense ISM.

\section{Summary}
\label{sec:summary}

Using idealized spherically symmetric models, we have studied galactic winds driven by CR streaming incorporating realistic radiative cooling. The inclusion of cooling is particularly important for studying winds from the warm ISM; cooling is comparatively less important for the hot ISM. The wind solutions found here exhibit distinctive features not present in winds neglecting radiative cooling or assuming that the gas is isothermal (which corresponds to the limit of extremely rapid cooling).

Near the base of the wind, where the wind speed is low, the density and temperature profiles can be roughly approximated as hydrostatic and arising from a balance between CR heating and cooling; see Figure \ref{fig:T_spike_approx}. This balance is, however, linearly unstable once the temperature exceeds $T\sim 1.75\times10^4\,$K \citep{Kempski_thermal_instability}. To study the effects of thermal instability, we carried out time-dependent numerical simulations of CR-driven winds using {\tt Athena++} with parameters appropriate for the warm ISM. Across a range of gravitational potentials, magnetic field strengths, base gas densities, and base CR pressures, we find that the winds broadly consist of 3 regimes: a nearly isothermal base in which heating and cooling balance, a region where CR heating dominates over cooling and expansion, and a region where the solution is nearly adiabatic. We find that the key parameter determining the properties of the wind and the spatial extent of these three regimes is the depth of the gravitational potential, which we parametrize by the potential's escape speed from $r = 0$ (see Figure \ref{fig:summary_profiles}).

When the escape speed is low, cooling remains strong throughout the wind, so thermal instability does not set in and the wind remains nearly isothermal. These nearly isothermal solutions are similar to the (fully) isothermal solutions in \citet{Quataert_streaming}. In contrast, at higher escape speeds, thermal instability causes large fluctuations in density and temperature at intermediate radii (see Figures \ref{fig:time_dependence} and \ref{fig:rho_variance}). Because the density decreases at larger radii, cooling becomes progressively less important relative to CR heating, and the thermal instability inevitably ``saturates'' with a sharp increase in temperature. This leads to a thermal gas pressure dominated wind at larger radii. Previous wind solutions with CR heating but neglecting cooling (e.g., \citealt{Ipavich1975,Everett2008}) effectively start from relatively high temperature ``base'' conditions, and accurately describe the structure of our solutions at larger radii where cooling is negligible. Our calculations show how these non-radiative models can be self-consistently extended deeper into a galaxy starting from physical conditions in the warm ISM.

Although our time-dependent solutions show some evidence for the acoustic instabilities studied in \cite{Begelman1994, Quataert_streaming}, and \cite{Tsung_Oh_Jiang_Staircase}, we find that thermal instability is by far the dominant source of large amplitude variability in our models; this is particularly true at intermediate escape speeds where cooling and heating remain comparable (but linearly unstable) for a range of radii. 

The asymptotic wind speed in our models scales approximately linearly with the escape speed, but is only at most $\sim 50\%$ of the escape speed. This low asymptotic speed implies that most of the CR energy supplied to the wind at the base is used to lift material out of the galaxy's gravitational potential. This energy balance argument can be used estimate the mass-loss rate, maximum temperature, and maximum outflow speed of the wind, as we demonstrate in \S\ref{sec:mdot}. The mass-loss rates we find can be comparable to or larger than the star formation rate in lower mass galaxies and they obey a roughly energy-like scaling of $\dot M \propto v_{\rm esc}^{-2}$. Our mass-loss rates are, however, on the low end of what is required to reconcile the galaxy stellar and halo mass functions. Because most of wind energy is lost escaping the gravitational potential of the galaxy, the asymptotic wind energy flux in our models is only $\sim 10\%$ of the input CR power, and thus $\sim 1\%$ of the input supernova power. These winds are thus inefficient in providing preventive feedback in the CGM.

Theoretically, the inclusion of cooling and CR heating in the dynamics of galactic winds leads to a unique critical point structure that defies textbook expectations (e.g. \citealt{Lamers1999}). In particular, the total sound speed of the gas-CR system  is imaginary when $v \ll v_A$ \citep{Ipavich1975}. Absent cooling, wind solutions only exist if the base velocity is large enough to avoid this unusual property of the CR hydrodynamic equations (e.g. \citealt{Ipavich1975, Everett2008}). With cooling, a wind with $v \ll v_A$ is possible, but is in a formal sense supersonic near its base (because $v^2 > 0 > v_d^2$, where $v_d$ is the critical point speed; see equations \ref{eq:v_profile}-\ref{eq:vn}).

Although the winds we find are time-dependent, the time-averaged wind profiles pass through two critical points matching the properties predicted by steady-state theory (see Figure \ref{fig:crit_pt_splitting}). The initial critical point occurs around where $v_d \sim 0$, i.e., where the CR sound speed becomes real. Formally this is analogous to a super-sonic to sub-sonic transition that is traditionally discarded as a possibility in wind models for being acausal. The second critical point we find is at larger radii and is the more conventional Parker-type critical point (see \S\ref{sec:critical} and \S\ref{sec:critical_sim}). The solutions we find thus have two critical points rather than the odd number traditionally expected. These are, to the best of our knowledge, the only wind solutions with these unusual properties, which are a consequence of both the imaginary CR sound speed and strong cooling.

A key observational signature of the winds found here is their slow acceleration to large radii. This leads to spatially extended emission and absorption lines from the optical to the UV (see Figure \ref{fig:luminosity_plots}). Up to $\sim 10\%$ of the wind's radiative luminosity is produced at radii larger than a few times the base radius, and is emitted over a wide range of $T\sim 10^4 - 10^{5.5}$\,K in all but the nearly isothermal models. This regime of the wind may directly correspond to the extraplanar diffuse ionized gas observed in many star-forming galaxies. Additionally, the variability due to thermal instability present across nearly all of our models strongly suggests that the gas develops a multiphase structure (as in \citealt{HuangDavisThermal}). An important direction for future work is to study the nonlinear outcome of thermal instability in multiple dimensions and its impact on both CR transport and the observational properties of galactic winds. Other generalizations of this work could involve including the bounding pressure of the CGM, which we neglect, including a more realistic disk-like geometry for the wind streamlines, and including photo-heating to develop a more realistic ISM model.

\section*{Data Availability}

The numerical simulation results used in this paper will be shared on request to the corresponding author.

\section*{Acknowledgements}

We thank Eve Ostriker and Navin Tsung for useful conversations. S.M. acknowledges support from the National Science Foundation Graduate Research Fellowship under Grant No. DGE-2039656. T.A.T. is supported in part by NASA \#80NSSC18K0526. This work was also supported in part by a Simons Investigator grant from the Simons Foundation and by NSF AST grant 2107872. The analysis presented in this article was performed in part on computational resources managed and supported by Princeton Research Computing, a consortium of groups including the Princeton Institute for Computational Science and Engineering (PICScie) and the Office of Information Technology's High Performance Computing Center and Visualization Laboratory at Princeton University.


\begin{thebibliography}{}
\makeatletter
\relax
\def\mn@urlcharsother{\let\do\@makeother \do\$\do\&\do\#\do\^\do\_\do\%\do\~}
\def\mn@doi{\begingroup\mn@urlcharsother \@ifnextchar [ {\mn@doi@}
  {\mn@doi@[]}}
\def\mn@doi@[#1]#2{\def\@tempa{#1}\ifx\@tempa\@empty \href
  {http://dx.doi.org/#2} {doi:#2}\else \href {http://dx.doi.org/#2} {#1}\fi
  \endgroup}
\def\mn@eprint#1#2{\mn@eprint@#1:#2::\@nil}
\def\mn@eprint@arXiv#1{\href {http://arxiv.org/abs/#1} {{\tt arXiv:#1}}}
\def\mn@eprint@dblp#1{\href {http://dblp.uni-trier.de/rec/bibtex/#1.xml}
  {dblp:#1}}
\def\mn@eprint@#1:#2:#3:#4\@nil{\def\@tempa {#1}\def\@tempb {#2}\def\@tempc
  {#3}\ifx \@tempc \@empty \let \@tempc \@tempb \let \@tempb \@tempa \fi \ifx
  \@tempb \@empty \def\@tempb {arXiv}\fi \@ifundefined
  {mn@eprint@\@tempb}{\@tempb:\@tempc}{\expandafter \expandafter \csname
  mn@eprint@\@tempb\endcsname \expandafter{\@tempc}}}

\bibitem[\protect\citeauthoryear{{Bai}}{{Bai}}{2022}]{Bai2022}
{Bai} X.-N.,  2022, \mn@doi [\apj] {10.3847/1538-4357/ac56e1}, \href
  {https://ui.adsabs.harvard.edu/abs/2022ApJ...928..112B} {928, 112}

\bibitem[\protect\citeauthoryear{{Bai}, {Ostriker}, {Plotnikov}  \&
  {Stone}}{{Bai} et~al.}{2019}]{Bai2019}
{Bai} X.-N.,  {Ostriker} E.~C.,  {Plotnikov} I.,   {Stone} J.~M.,  2019,
  \mn@doi [\apj] {10.3847/1538-4357/ab1648}, \href
  {https://ui.adsabs.harvard.edu/abs/2019ApJ...876...60B} {876, 60}

\bibitem[\protect\citeauthoryear{{Begelman} \& {Zweibel}}{{Begelman} \&
  {Zweibel}}{1994}]{Begelman1994}
{Begelman} M.~C.,  {Zweibel} E.~G.,  1994, \mn@doi [\apj] {10.1086/174519},
  \href {https://ui.adsabs.harvard.edu/abs/1994ApJ...431..689B} {431, 689}

\bibitem[\protect\citeauthoryear{{Blasi}, {Amato}  \& {Serpico}}{{Blasi}
  et~al.}{2012}]{Blasi2012}
{Blasi} P.,  {Amato} E.,   {Serpico} P.~D.,  2012, \mn@doi [\prl]
  {10.1103/PhysRevLett.109.061101}, \href
  {https://ui.adsabs.harvard.edu/abs/2012PhRvL.109f1101B} {109, 061101}

\bibitem[\protect\citeauthoryear{{Booth}, {Agertz}, {Kravtsov}  \&
  {Gnedin}}{{Booth} et~al.}{2013}]{Booth2013}
{Booth} C.~M.,  {Agertz} O.,  {Kravtsov} A.~V.,   {Gnedin} N.~Y.,  2013,
  \mn@doi [\apjl] {10.1088/2041-8205/777/1/L16}, \href
  {https://ui.adsabs.harvard.edu/abs/2013ApJ...777L..16B} {777, L16}

\bibitem[\protect\citeauthoryear{{Boulares} \& {Cox}}{{Boulares} \&
  {Cox}}{1990}]{Boulares1990}
{Boulares} A.,  {Cox} D.~P.,  1990, \mn@doi [\apj] {10.1086/169509}, \href
  {https://ui.adsabs.harvard.edu/abs/1990ApJ...365..544B} {365, 544}

\bibitem[\protect\citeauthoryear{{Breitschwerdt}, {McKenzie}  \&
  {Voelk}}{{Breitschwerdt} et~al.}{1991}]{Breitschwerdt1991}
{Breitschwerdt} D.,  {McKenzie} J.~F.,   {Voelk} H.~J.,  1991, \aap, \href
  {https://ui.adsabs.harvard.edu/abs/1991A&A...245...79B} {245, 79}

\bibitem[\protect\citeauthoryear{{Chan}, {Kere{\v{s}}}, {Hopkins}, {Quataert},
  {Su}, {Hayward}  \& {Faucher-Gigu{\`e}re}}{{Chan} et~al.}{2019}]{Chan2019}
{Chan} T.~K.,  {Kere{\v{s}}} D.,  {Hopkins} P.~F.,  {Quataert} E.,  {Su} K.~Y.,
   {Hayward} C.~C.,   {Faucher-Gigu{\`e}re} C.~A.,  2019, \mn@doi [\mnras]
  {10.1093/mnras/stz1895}, \href
  {https://ui.adsabs.harvard.edu/abs/2019MNRAS.488.3716C} {488, 3716}

\bibitem[\protect\citeauthoryear{{Draine}}{{Draine}}{2011}]{Draine}
{Draine} B.~T.,  2011, {Physics of the Interstellar and Intergalactic Medium}.
Princeton Univ. Press, Princeton

\bibitem[\protect\citeauthoryear{{Everett}, {Zweibel}, {Benjamin}, {McCammon},
  {Rocks}  \& {Gallagher}}{{Everett} et~al.}{2008}]{Everett2008}
{Everett} J.~E.,  {Zweibel} E.~G.,  {Benjamin} R.~A.,  {McCammon} D.,  {Rocks}
  L.,   {Gallagher} John~S. I.,  2008, \mn@doi [\apj] {10.1086/524766}, \href
  {https://ui.adsabs.harvard.edu/abs/2008ApJ...674..258E} {674, 258}

\bibitem[\protect\citeauthoryear{{Girichidis} et~al.,}{{Girichidis}
  et~al.}{2016}]{Girichidi2016}
{Girichidis} P.,  et~al., 2016, \mn@doi [\apjl] {10.3847/2041-8205/816/2/L19},
  \href {https://ui.adsabs.harvard.edu/abs/2016ApJ...816L..19G} {816, L19}

\bibitem[\protect\citeauthoryear{{Guo} \& {Oh}}{{Guo} \& {Oh}}{2008}]{Guo2008}
{Guo} F.,  {Oh} S.~P.,  2008, \mn@doi [\mnras]
  {10.1111/j.1365-2966.2007.12692.x}, \href
  {https://ui.adsabs.harvard.edu/abs/2008MNRAS.384..251G} {384, 251}

\bibitem[\protect\citeauthoryear{{Heckman}, {Alexandroff}, {Borthakur},
  {Overzier}  \& {Leitherer}}{{Heckman} et~al.}{2015}]{Heckman2015}
{Heckman} T.~M.,  {Alexandroff} R.~M.,  {Borthakur} S.,  {Overzier} R.,
  {Leitherer} C.,  2015, \mn@doi [\apj] {10.1088/0004-637X/809/2/147}, \href
  {https://ui.adsabs.harvard.edu/abs/2015ApJ...809..147H} {809, 147}

\bibitem[\protect\citeauthoryear{{Hernquist}}{{Hernquist}}{1990}]{Hernquist_potential}
{Hernquist} L.,  1990, \mn@doi [\apj] {10.1086/168845}, \href
  {https://ui.adsabs.harvard.edu/abs/1990ApJ...356..359H} {356, 359}

\bibitem[\protect\citeauthoryear{{Hopkins}, {Squire}, {Butsky}  \&
  {Ji}}{{Hopkins} et~al.}{2022}]{Hopkins2022}
{Hopkins} P.~F.,  {Squire} J.,  {Butsky} I.~S.,   {Ji} S.,  2022, \mn@doi
  [\mnras] {10.1093/mnras/stac2909}, \href
  {https://ui.adsabs.harvard.edu/abs/2022MNRAS.517.5413H} {517, 5413}

\bibitem[\protect\citeauthoryear{{Huang} \& {Davis}}{{Huang} \&
  {Davis}}{2022}]{HuangDavisLaunching}
{Huang} X.,  {Davis} S.~W.,  2022, \mn@doi [\mnras] {10.1093/mnras/stac059},
  \href {https://ui.adsabs.harvard.edu/abs/2022MNRAS.511.5125H} {511, 5125}

\bibitem[\protect\citeauthoryear{{Huang}, {Jiang}  \& {Davis}}{{Huang}
  et~al.}{2022}]{HuangDavisThermal}
{Huang} X.,  {Jiang} Y.-f.,   {Davis} S.~W.,  2022, \mn@doi [\apj]
  {10.3847/1538-4357/ac69dc}, \href
  {https://ui.adsabs.harvard.edu/abs/2022ApJ...931..140H} {931, 140}

\bibitem[\protect\citeauthoryear{{Ipavich}}{{Ipavich}}{1975}]{Ipavich1975}
{Ipavich} F.~M.,  1975, \mn@doi [\apj] {10.1086/153397}, \href
  {https://ui.adsabs.harvard.edu/abs/1975ApJ...196..107I} {196, 107}

\bibitem[\protect\citeauthoryear{{Jacob} \& {Pfrommer}}{{Jacob} \&
  {Pfrommer}}{2017}]{Svenja2017}
{Jacob} S.,  {Pfrommer} C.,  2017, \mn@doi [\mnras] {10.1093/mnras/stx131},
  \href {https://ui.adsabs.harvard.edu/abs/2017MNRAS.467.1449J} {467, 1449}

\bibitem[\protect\citeauthoryear{{Jacob}, {Pakmor}, {Simpson}, {Springel}  \&
  {Pfrommer}}{{Jacob} et~al.}{2018}]{Jacob2018}
{Jacob} S.,  {Pakmor} R.,  {Simpson} C.~M.,  {Springel} V.,   {Pfrommer} C.,
  2018, \mn@doi [\mnras] {10.1093/mnras/stx3221}, \href
  {https://ui.adsabs.harvard.edu/abs/2018MNRAS.475..570J} {475, 570}

\bibitem[\protect\citeauthoryear{{Ji}, {Oh}  \& {McCourt}}{{Ji}
  et~al.}{2018}]{Ji_cooling}
{Ji} S.,  {Oh} S.~P.,   {McCourt} M.,  2018, \mn@doi [\mnras]
  {10.1093/mnras/sty293}, \href
  {https://ui.adsabs.harvard.edu/abs/2018MNRAS.476..852J} {476, 852}

\bibitem[\protect\citeauthoryear{{Ji} et~al.,}{{Ji} et~al.}{2020}]{Ji2020}
{Ji} S.,  et~al., 2020, \mn@doi [\mnras] {10.1093/mnras/staa1849}, \href
  {https://ui.adsabs.harvard.edu/abs/2020MNRAS.496.4221J} {496, 4221}

\bibitem[\protect\citeauthoryear{{Jiang} \& {Oh}}{{Jiang} \&
  {Oh}}{2018}]{Jiang_Oh_Transport}
{Jiang} Y.-F.,  {Oh} S.~P.,  2018, \mn@doi [\apj] {10.3847/1538-4357/aaa6ce},
  \href {https://ui.adsabs.harvard.edu/abs/2018ApJ...854....5J} {854, 5}

\bibitem[\protect\citeauthoryear{{Kempski} \& {Quataert}}{{Kempski} \&
  {Quataert}}{2020}]{Kempski_thermal_instability}
{Kempski} P.,  {Quataert} E.,  2020, \mn@doi [\mnras] {10.1093/mnras/staa385},
  \href {https://ui.adsabs.harvard.edu/abs/2020MNRAS.493.1801K} {493, 1801}

\bibitem[\protect\citeauthoryear{{Kempski} \& {Quataert}}{{Kempski} \&
  {Quataert}}{2022}]{Kempski2022}
{Kempski} P.,  {Quataert} E.,  2022, \mn@doi [\mnras] {10.1093/mnras/stac1240},
  \href {https://ui.adsabs.harvard.edu/abs/2022MNRAS.514..657K} {514, 657}

\bibitem[\protect\citeauthoryear{{Kulsrud} \& {Pearce}}{{Kulsrud} \&
  {Pearce}}{1969}]{Kulsrud1969}
{Kulsrud} R.,  {Pearce} W.~P.,  1969, \mn@doi [\apj] {10.1086/149981}, \href
  {https://ui.adsabs.harvard.edu/abs/1969ApJ...156..445K} {156, 445}

\bibitem[\protect\citeauthoryear{{Lacki}, {Thompson}  \& {Quataert}}{{Lacki}
  et~al.}{2010}]{Lacki2010}
{Lacki} B.~C.,  {Thompson} T.~A.,   {Quataert} E.,  2010, \mn@doi [\apj]
  {10.1088/0004-637X/717/1/110.48550/arXiv.0907.4161}, \href
  {https://ui.adsabs.harvard.edu/abs/2010ApJ...717....1L} {717, 1}

\bibitem[\protect\citeauthoryear{{Lamers} \& {Cassinelli}}{{Lamers} \&
  {Cassinelli}}{1999}]{Lamers1999}
{Lamers} H. J.~G.~L.~M.,  {Cassinelli} J.~P.,  1999, {Introduction to Stellar
  Winds}

\bibitem[\protect\citeauthoryear{{Lu} et~al.,}{{Lu} et~al.}{2023}]{Lu2023_eDIG}
{Lu} L.-Y.,  et~al., 2023, \mn@doi [\mnras] {10.1093/mnras/stad006}, \href
  {https://ui.adsabs.harvard.edu/abs/2023MNRAS.519.6098L} {519, 6098}

\bibitem[\protect\citeauthoryear{{Mao} \& {Ostriker}}{{Mao} \&
  {Ostriker}}{2018}]{Mao_Ostriker}
{Mao} S.~A.,  {Ostriker} E.~C.,  2018, \mn@doi [\apj]
  {10.3847/1538-4357/aaa88e}, \href
  {https://ui.adsabs.harvard.edu/abs/2018ApJ...854...89M} {854, 89}

\bibitem[\protect\citeauthoryear{{Marasco} et~al.,}{{Marasco}
  et~al.}{2022}]{Marasco_ShakenBlown}
{Marasco} A.,  et~al., 2022, \mn@doi [arXiv e-prints]
  {10.48550/arXiv.2209.02726}, \href
  {https://ui.adsabs.harvard.edu/abs/2022arXiv220902726M} {p. arXiv:2209.02726}

\bibitem[\protect\citeauthoryear{{McQuinn}, {van Zee}  \& {Skillman}}{{McQuinn}
  et~al.}{2019}]{McQuinn19}
{McQuinn} K. B.~W.,  {van Zee} L.,   {Skillman} E.~D.,  2019, \mn@doi [\apj]
  {10.3847/1538-4357/ab4c37}, \href
  {https://ui.adsabs.harvard.edu/abs/2019ApJ...886...74M} {886, 74}

\bibitem[\protect\citeauthoryear{{Parker}}{{Parker}}{1958}]{Parker_wind}
{Parker} E.~N.,  1958, \mn@doi [\apj] {10.1086/146579}, \href
  {https://ui.adsabs.harvard.edu/abs/1958ApJ...128..664P} {128, 664}

\bibitem[\protect\citeauthoryear{{Pfrommer}, {Pakmor}, {Schaal}, {Simpson}  \&
  {Springel}}{{Pfrommer} et~al.}{2017}]{Pfrommer2017}
{Pfrommer} C.,  {Pakmor} R.,  {Schaal} K.,  {Simpson} C.~M.,   {Springel} V.,
  2017, \mn@doi [\mnras] {10.1093/mnras/stw2941}, \href
  {http://adsabs.harvard.edu/abs/2017MNRAS.465.4500P} {465, 4500}

\bibitem[\protect\citeauthoryear{{Quataert}, {Jiang}  \& {Thompson}}{{Quataert}
  et~al.}{2022a}]{Quataert_streaming}
{Quataert} E.,  {Jiang} F.,   {Thompson} T.~A.,  2022a, \mn@doi [\mnras]
  {10.1093/mnras/stab3274}, \href
  {https://ui.adsabs.harvard.edu/abs/2022MNRAS.510..920Q} {510, 920}

\bibitem[\protect\citeauthoryear{{Quataert}, {Thompson}  \& {Jiang}}{{Quataert}
  et~al.}{2022b}]{Quataert_diffusion}
{Quataert} E.,  {Thompson} T.~A.,   {Jiang} Y.-F.,  2022b, \mn@doi [\mnras]
  {10.1093/mnras/stab3273}, \href
  {https://ui.adsabs.harvard.edu/abs/2022MNRAS.510.1184Q} {510, 1184}

\bibitem[\protect\citeauthoryear{{Rathjen} et~al.,}{{Rathjen}
  et~al.}{2021}]{Rathjen2021}
{Rathjen} T.-E.,  et~al., 2021, \mn@doi [\mnras] {10.1093/mnras/stab900}, \href
  {https://ui.adsabs.harvard.edu/abs/2021MNRAS.504.1039R} {504, 1039}

\bibitem[\protect\citeauthoryear{{Rautio}, {Watkins}, {Comer{\'o}n}, {Salo},
  {D{\'\i}az-Garc{\'\i}a}  \& {Janz}}{{Rautio} et~al.}{2022}]{Rautio_2022}
{Rautio} R.~P.~V.,  {Watkins} A.~E.,  {Comer{\'o}n} S.,  {Salo} H.,
  {D{\'\i}az-Garc{\'\i}a} S.,   {Janz} J.,  2022, \mn@doi [\aap]
  {10.1051/0004-6361/202142440}, \href
  {https://ui.adsabs.harvard.edu/abs/2022A&A...659A.153R} {659, A153}

\bibitem[\protect\citeauthoryear{{Ruszkowski}, {Yang}  \&
  {Zweibel}}{{Ruszkowski} et~al.}{2017}]{Ruszkowski2017}
{Ruszkowski} M.,  {Yang} H. Y.~K.,   {Zweibel} E.,  2017, \mn@doi [\apj]
  {10.3847/1538-4357/834/2/208}, \href
  {https://ui.adsabs.harvard.edu/#abs/2017ApJ...834..208R} {834, 208}

\bibitem[\protect\citeauthoryear{{Simpson}, {Pakmor}, {Pfrommer}, {Glover}  \&
  {Smith}}{{Simpson} et~al.}{2023}]{Simpson2023}
{Simpson} C.~M.,  {Pakmor} R.,  {Pfrommer} C.,  {Glover} S. C.~O.,   {Smith}
  R.,  2023, \mn@doi [\mnras] {10.1093/mnras/stac3601}, \href
  {https://ui.adsabs.harvard.edu/abs/2023MNRAS.520.4621S} {520, 4621}

\bibitem[\protect\citeauthoryear{{Skilling}}{{Skilling}}{1971}]{Skilling1971}
{Skilling} J.,  1971, \mn@doi [\apj] {10.1086/151210}, \href
  {https://ui.adsabs.harvard.edu/abs/1971ApJ...170..265S} {170, 265}

\bibitem[\protect\citeauthoryear{{Socrates}, {Davis}  \&
  {Ramirez-Ruiz}}{{Socrates} et~al.}{2008}]{SocratesDavisRamirezRuiz}
{Socrates} A.,  {Davis} S.~W.,   {Ramirez-Ruiz} E.,  2008, \mn@doi [\apj]
  {10.1086/590046}, \href
  {https://ui.adsabs.harvard.edu/abs/2008ApJ...687..202S} {687, 202}

\bibitem[\protect\citeauthoryear{{Somerville} \& {Dav{\'e}}}{{Somerville} \&
  {Dav{\'e}}}{2015}]{Somverville2015}
{Somerville} R.~S.,  {Dav{\'e}} R.,  2015, \mn@doi [\araa]
  {10.1146/annurev-astro-082812-140951}, \href
  {https://ui.adsabs.harvard.edu/abs/2015ARA&A..53...51S} {53, 51}

\bibitem[\protect\citeauthoryear{{Stone}, {Tomida}, {White}  \&
  {Felker}}{{Stone} et~al.}{2020}]{Stone_athena}
{Stone} J.~M.,  {Tomida} K.,  {White} C.~J.,   {Felker} K.~G.,  2020, \mn@doi
  [\apjs] {10.3847/1538-4365/ab929b}, \href
  {https://ui.adsabs.harvard.edu/abs/2020ApJS..249....4S} {249, 4}

\bibitem[\protect\citeauthoryear{{Thomas}, {Pfrommer}  \& {Pakmor}}{{Thomas}
  et~al.}{2021}]{Thomas2021}
{Thomas} T.,  {Pfrommer} C.,   {Pakmor} R.,  2021, \mn@doi [\mnras]
  {10.1093/mnras/stab397}, \href
  {https://ui.adsabs.harvard.edu/abs/2021MNRAS.503.2242T} {503, 2242}

\bibitem[\protect\citeauthoryear{{Tsung}, {Oh}  \& {Jiang}}{{Tsung}
  et~al.}{2022}]{Tsung_Oh_Jiang_Staircase}
{Tsung} T. H.~N.,  {Oh} S.~P.,   {Jiang} Y.-F.,  2022, \mn@doi [\mnras]
  {10.1093/mnras/stac1123}, \href
  {https://ui.adsabs.harvard.edu/abs/2022MNRAS.513.4464T} {513, 4464}

\bibitem[\protect\citeauthoryear{{Uhlig}, {Pfrommer}, {Sharma}, {Nath},
  {En{\ss}lin}  \& {Springel}}{{Uhlig} et~al.}{2012}]{Uhligh2012}
{Uhlig} M.,  {Pfrommer} C.,  {Sharma} M.,  {Nath} B.~B.,  {En{\ss}lin} T.~A.,
  {Springel} V.,  2012, \mn@doi [\mnras] {10.1111/j.1365-2966.2012.21045.x},
  \href {https://ui.adsabs.harvard.edu/abs/2012MNRAS.423.2374U} {423, 2374}

\bibitem[\protect\citeauthoryear{{Veilleux}, {Cecil}  \&
  {Bland-Hawthorn}}{{Veilleux} et~al.}{2005}]{Veilleux2005}
{Veilleux} S.,  {Cecil} G.,   {Bland-Hawthorn} J.,  2005, \mn@doi [\araa]
  {10.1146/annurev.astro.43.072103.150610}, \href
  {https://ui.adsabs.harvard.edu/abs/2005ARA&A..43..769V} {43, 769}

\bibitem[\protect\citeauthoryear{{Weiner} et~al.,}{{Weiner}
  et~al.}{2009}]{Weiner2009}
{Weiner} B.~J.,  et~al., 2009, \mn@doi [\apj] {10.1088/0004-637X/692/1/187},
  \href {https://ui.adsabs.harvard.edu/abs/2009ApJ...692..187W} {692, 187}

\bibitem[\protect\citeauthoryear{{Wentzel}}{{Wentzel}}{1971}]{Wentzel1971}
{Wentzel} D.~G.,  1971, \mn@doi [\apj] {10.1086/150794}, \href
  {https://ui.adsabs.harvard.edu/abs/1971ApJ...163..503W} {163, 503}

\bibitem[\protect\citeauthoryear{{Wiener}, {Zweibel}  \& {Oh}}{{Wiener}
  et~al.}{2013}]{Wiener_Zweibel_2013}
{Wiener} J.,  {Zweibel} E.~G.,   {Oh} S.~P.,  2013, \mn@doi [\apj]
  {10.1088/0004-637X/767/1/87}, \href
  {https://ui.adsabs.harvard.edu/abs/2013ApJ...767...87W} {767, 87}

\bibitem[\protect\citeauthoryear{{Wiener}, {Oh}  \& {Zweibel}}{{Wiener}
  et~al.}{2017}]{Wiener2017}
{Wiener} J.,  {Oh} S.~P.,   {Zweibel} E.~G.,  2017, \mn@doi [\mnras]
  {10.1093/mnras/stx10910.48550/arXiv.1610.02041}, \href
  {https://ui.adsabs.harvard.edu/abs/2017MNRAS.467..646W} {467, 646}

\bibitem[\protect\citeauthoryear{{Wiener}, {Zweibel}  \& {Oh}}{{Wiener}
  et~al.}{2018}]{Wiener2018}
{Wiener} J.,  {Zweibel} E.~G.,   {Oh} S.~P.,  2018, \mn@doi [\mnras]
  {10.1093/mnras/stx2603}, \href
  {https://ui.adsabs.harvard.edu/abs/2018MNRAS.473.3095W} {473, 3095}

\bibitem[\protect\citeauthoryear{{Wiener}, {Zweibel}  \& {Ruszkowski}}{{Wiener}
  et~al.}{2019}]{Wiener2019}
{Wiener} J.,  {Zweibel} E.~G.,   {Ruszkowski} M.,  2019, \mn@doi [\mnras]
  {10.1093/mnras/stz200710.48550/arXiv.1903.01471}, \href
  {https://ui.adsabs.harvard.edu/abs/2019MNRAS.489..205W} {489, 205}

\bibitem[\protect\citeauthoryear{{Xu} et~al.,}{{Xu} et~al.}{2023}]{Xu2023}
{Xu} X.,  et~al., 2023, \mn@doi [\apj] {10.3847/1538-4357/acbf46}, \href
  {https://ui.adsabs.harvard.edu/abs/2023ApJ...948...28X} {948, 28}

\makeatother
\end{thebibliography}

\bsp
\label{lastpage}

\end{document}